\documentclass[utf8]{frontiersFPHY}
\usepackage{graphicx}

\usepackage{url,hyperref,microtype,subcaption}
\usepackage[onehalfspacing]{setspace}

\def\firstAuthorLast{A Rios}
\def\Authors{Arnau Rios}

\def\Address{Department of Physics, Faculty of Engineering and Physical Sciences, University of Surrey,
Guildford, Surrey GU2 7XH, United Kingdom}

\def\corrEmail{a.rios@surrey.ac.uk}
\def\corrAuthor{Corresponding Author}

\begin{document}
\onecolumn
\firstpage{1}

\title[Green's functions techniques for extended nuclear systems]{Green's functions techniques for extended nuclear systems} 

\author[\firstAuthorLast]{\Authors} 
\address{} 
\correspondance{} 

\extraAuth{}

\maketitle

\begin{abstract}
I review the application of self-consistent Green's functions methods to study the properties of infinite nuclear systems. Improvements over the last decade, including the consistent treatment of three-nucleon forces and the development of extrapolation methods from finite to zero temperature, have allowed for realistic predictions of the equation of state of infinite symmetric, asymmetric and neutron matter based on chiral interactions. Microscopic properties, like momentum distributions or spectral functions, are also accessible. Using an indicative set of results based on a subset of chiral interactions, I summarise here the first-principles description of infinite nuclear system provided by Green's functions techniques, in the context of several issues of relevance for nuclear theory including, but not limited to, the role of short-range correlations in nuclear systems, nuclear phase transitions and the isospin dependence of nuclear observables.
\end{abstract}

\section{Introduction}

The recent discoveries of neutron-star binaries GW170817 \cite{GW170817discovery} and GW190425 \cite{GW190425discovery} are formidable feats in gravitational-wave (GW) and multimessenger astronomy. These events provide unique insight into a great variety of astrophysical questions, from stellar physics to cosmology. In dense matter physics, the analysis of multimessenger data is providing a unique way forward in our understanding of matter in its most neutron-rich form. Gravitational waves from the inspiral carry imprints of the equation of state (EoS) of neutron-star matter \cite{GW170817MR1}. The initial analysis of the source GW170817 has already provided significant constraints on the masses, radii and EoSs of neutron stars \cite{GW170817MR2}. The expectation is that, as more of these GW sources are detected in the next few years, they will provide more and more accurate observational constraints. Simultaneous developments in X-ray burst observations are also providing equally important and independent constraints on the EoS and the physical properties of neutron stars (NSs) \cite{Raaijmakers2020}. 

Theoretical nuclear physics tools provide relevant input for dense matter physics. Among others, there have been several significant efforts trying to put constraints on the symmetry energy \cite{Tsang2012,Drischler2020} or the EoS itself \cite{Danielewicz2002}. The data in the new generation of astrophysical observations will tighten these constraints, and may even allow for further relevant conclusions on macroscopic observables, like the EoS \cite{Tews2020}. Connecting these bulk observable constraints to the strong interaction among neutrons requires significant efforts on the nuclear theory side. There is a stringent need for the developments of realistic predictions with systematic errors based on our understanding of the physics of dense systems.  This necessarily requires the use of \emph{ab initio} nuclear theory techniques that can provide a solid, quantifiable link between macro- and micro-physics \cite{Baym2018,Lim2018}. For the case of GW inspirals, this link also demands for an understanding of dense matter properties not only in the zero-temperature domain, but at temperature scales which are significant enough to modify some nuclear properties (e.g. temperatures in the range of $10$s of MeV) \cite{Constantinou2014,Constantinou2015,Baiotti2017,Duez2018,Carbone2019}. There is a similar need for microphysics observations in the case of core-collapse supernovae, where the formation process of proto-NSs \cite{Pons1999} and the subsequent development of a neutrino-heated shock wave require an understanding of neutrino emission from compact objects \cite{MartinezPinedo2012,Rrapaj2015}. 

The accuracy of theoretical nuclear physics predictions is hampered by a variety of different factors. First, the strong interaction among nucleons is difficult to pin down. Scattering experiments provide extremely useful insight at the two-body level, but three-neutron forces remain relatively unconstrained \cite{Gandolfi2017}. For neutrons, four-body and higher level forces are expected to play a small role for the EoS \cite{Tews2013}. Chiral effective field theory ($\chi$EFT) \cite{Epelbaum2009,Machleidt2011} and Lattice QCD \cite{Ishii2007,Hatsuda2018} efforts  in the next decade will be crucial to provide clear limits on these terms. As a first approximation, one can estimate the uncertainty in the strong force by considering different input hamiltonians that fit scattering phase-shifts and few body data in in infinite matter simulations \cite{Drischler2016b,Drischler2019}. 

In addition, even if the interaction is perfectly understood, one still needs to solve a many-body multicomponent\footnote{Multicomponent in the sense that it requires both neutrons and protons.} problem to link the NN force to the macroscopic properties related to the EoS. While several methods have been devised to attack this problem, a clear-cut  solution is difficult, if not impossible. The treatment of  many-body correlations differs depending on the method that is used to describe dense neutron-rich systems \cite{Baldo2012,Piarulli2020}. In particular, the structure of the strong force itself - particularly its short-range core - may preclude low-order perturbative treatments. Self-consistent Green's (SCGF) functions techniques can be used to obtain both perturbative and non-perturbative results from these interactions of relevance for nuclear physics \cite{Dickhoff2004,Dickhoff_book,Barbieri2017}. This approach offers the promise to simultaneously tackle issues that are relevant for neutron-star astrophysics, while providing insight into problems that are appropriate for nuclear structure (particularly in terms of short-range correlations). 

My aim with this review is to provide a summary of the recent advances of the SCGF treatment of infinite nuclear systems. I will only explore the uncertainty associated to the hamiltonian by looking at calculations performed with 3 different chiral interactions. This represents a small subset of possible forces, and is by no means representative of the overall uncertainty due to the strong interaction in infinite systems. Even then, these interactions predict relatively different micro- and macroscopic properties below, around and above saturation density. 
A benchmark analysis with different methods, while necessary to fully understand the limitations of nuclear theory, lies beyond the scope of this contribution. I will instead provide a qualitative, and somewhat limited discussion of the many-body scheme dependence of the results, by providing comparisons to many-body perturbation theory results based on similar hamiltonians. 

I start this review with a brief summary of the formal tools associated to the SCGF formalism in the following Section. I note that there are monographs \cite{Dickhoff2004,Barbieri2017} and books \cite{Dickhoff_book,Economou_book,vanLeeuwen_book} that provide many more details, and I refer the interested reader to those. Section~\ref{sec:micro} is devoted to a discussion of recent numerical results that concern single-particle properties in dense matter. These include spectral functions, self-energies and momentum distributions. In Section~\ref{sec:macro}, the discussion turns into the macrophysics of infinite nuclear systems, with a particular emphasis on the EoS. I finally provide some conclusions and outlook in Sec.~\ref{sec:conclusions}.

\section{Self-Consistent Green's Functions Approach}

\subsection{Formalism}

Many-body Green's functions are a natural starting point in the discussion of quantum many-body systems \cite{Abrikosov1965,Dickhoff_book}. Green's functions (or propagators) arise naturally in diagrammatic treatments, and provide a description of several real or virtual excitation processes within the many-body system. Formally, these propagators take the form of time-ordered correlation functions associated to the creation and annihilation of excitations and can be expressed in terms of Feynman diagrams \cite{Mattuck_book,Fetter_book}. Broadly speaking, Green's functions techniques attempt to sum series of the most physically relevant diagrams to describe the system under consideration. For instance, in the case of thermodynamics of infinite correlated systems, one often resorts to an expansion that involves renormalized interactions treating the short-range core non-perturbatively via infinitely repeated scattering processes \cite{Muther2000}. When it comes to the response of finite systems, one instead considers diagrams that describe relevant excitations within the system  \cite{Dickhoff2004}. This allows SCGF practitioners to look at many-body systems from different perspectives, providing a characterisation of the system that includes micro- and macroscopic properties.

In SCGFs methods, propagators in all internal diagrams are consistently dressed. In other words, renormalisation effects are introduced from the onset in the formalism - and hence the monicker "self-consistent". At a diagrammatic level, this means that only skeleton (or 1-particle-irreducible) diagrams are required for single-particle self-energies. In practice, one pays the price of having to run simulations iteratively for a given system which, in infinite matter, translates into a fixed density (or temperature) \cite{Dickhoff_book}. A key advantage of SCGF methods is the direct access to the propagators themselves. These contain relevant information for the system's microphysics, in terms of spectroscopy and strength distributions \cite{Dickhoff2004}. Moreover, propagators can also be used to compute bulk properties, including total energies, thermodynamical properties and pressures. Within a single SCGF simulation, one therefore obtains a significantly broad scope in terms of physical information. 
This goes beyond the reach of other many-body approaches which are mostly tailored to address the energy of the system, either by minimising it using quantum Monte-Carlo \cite{Gandolfi2015,Piarulli2020} techniques, or addressing it it by means of diagrams, like the Brueckner-Hartree-Fock (BHF) approach \cite{Baldo2012}. 

Another strategic advantage of SCGF formalisms are their generality. Within the very same formalism, one can tackle both finite and infinite systems - see Ref.~\cite{Soma2020} for a contribution in this volume that specifically discusses recent applications of SCGF techniques to finite nuclei. More importantly in the context of astrophysics, SCGF methods can be formulated consistently at finite temperature \cite{Abrikosov1965,Dickhoff_book}. In fact, fully-fledged infinite nuclear matter studies are usually performed at finite temperature \cite{Bozek2003,Rios2009b,Soma2009}. For sufficiently large temperatures (typically $T \gtrapprox 2-5$ MeV), this treatment avoids the complications associated to nuclear superfluidity \cite{Bozek1999,Bozek2003} that plagued  early attempts of SCGF calculations in the 1990s \cite{Polls1995}. Furthermore, a finite temperature formulation allows for access to astrophysically-relevant temperature dependences of observables \cite{Carbone2019}, and provides insight into liquid-gas, thermally-induced phase transitions in dense systems \cite{Rios2008,Carbone2018}. It is now possible to use these non-superfluid finite temperature calculations to perform meaningful extrapolations to zero temperatures for both microscopic \cite{Ding2016} and bulk properties \cite{Carbone2020}, as we shall see in the following. 

A key quantity in the SCGF formalism is the so-called spectral function, $\mathcal{A}_k(\omega)$. This characterises fully the one-body propagator, $\mathcal{G}$, which is generally defined as a a time-ordered product of a creation and a destruction operator \cite{Dickhoff_book,vanLeeuwen_book}. Upon Fourier transforming into a complex energy variable, $z$, the spectral function and the one-body propagator are related via a spectral decomposition,
\begin{align}
\mathcal{G}_k(z) = \int \frac{\text{d} \omega}{2 \pi} \frac{\mathcal{A}_k(\omega)}{z-\omega} \, .
\end{align}
Different time orderings give rise to different components of the propagator, including the retarded and advanced components that can be addressed by taking $z = \omega \pm i \eta$, with $\eta$ very small \cite{Economou_book}. All time-ordered components can be accessed through the spectral function, $\mathcal{A}_k(\omega)$, which also has the advantage of having a probabilistic interpretation. 
The spectral function provides the probability distribution associated to either extracting or attaching a particle with well-defined momentum, $k$, and energy, $\omega$, to the infinite system. In the finite-temperature Matsubara formalism, the Lehmann representation of $\mathcal{A}_k(\omega)$ takes into account  not only the 
thermal excitations on the ground state of the system, but also the presence of excited states that are populated according to a thermal distribution, 
\begin{align}
\mathcal{A}_k(\omega) &= 
2 \pi \sum_{n,m} \frac{e^{-\beta (E_n - \mu N_n)}}{Z} | \langle m | \hat a_{k}^{\dagger} | n \rangle |^2 \,
\delta [\omega - (E_m - E_n)]  \nonumber \\
&+ 2 \pi \sum_{n,m} \frac{e^{-\beta (E_n - \mu N_n)}}{Z}| \langle m | \hat a_{k}^{} | n \rangle |^2 \,
\delta [\omega - (E_n - E_m)] \, .
 \label{eq:sf_spectral}  
\end{align}
In this equation, $\hat a_k$ ($\hat a_k^\dagger$) are destruction (creation) operators of single-particle states with well defined momentum, $k$. A many-body system with $N_n$ particles can have different $E_n$ many-body hamiltonian eigenstates, labelled with quantum number $n$. $Z=\sum_n e^{-\beta (E_n - \mu N_n)}$ is the corresponding partition function and, in a grand-canonical setting, the system is described in terms of a given temperature, $T=1/\beta$, and chemical potential, $\mu$. 

As a well-defined probability distribution, the spectral function is normalized, $\int \text{d} \omega / 2 \pi \mathcal{A}_k (\omega) = 1$. When this integral is weighted by the corresponding thermal population of hole states, one obtains the momentum distribution of the system, 
\begin{align}
n_k =  \int \frac{\text{d} \omega}{2 \pi}  \mathcal{A}_k(\omega) f(\omega)  \, .
 \label{eq:momdis}  
\end{align}
In turn, even though the formulation of the problem is grand-canonical in nature, the momentum distribution can be used to fix the density of the system by requiring that the chemical potential in the Fermi-Dirac term $f(\omega)=\left[ 1 + \exp[ (\omega - \mu) / T ]\right]^{-1}$ is such that the normalisation condition
\begin{align}
\rho = \nu \int  \frac{\text{d}^3 k}{(2 \pi)^3} n_k 
 \label{eq:dens}  
\end{align}
 is fulfilled. Here, $\nu=2$ represents the degeneracy in spin in neutron or asymmetric matter, and $\nu=4$ is the degeneracy in spin-isospin for symmetric matter.
 
The Dyson equation, $[ z  - k^2/2m  - \Sigma_k(z)] \mathcal{G}_k(z) = 1$ provides a connection between the one-body propagator, $\mathcal{G}$, and the self-energy, $\Sigma$. The latter is traditionally decomposed into a real and an imaginary part, and the spectral function is then connected to the retarded self-energy components by the Lorentzian-like equation:
\begin{align}
\mathcal{A}_k(\omega) =  \frac{-2 \text{Im} \Sigma_k(\omega)}{\big[ \omega - \frac{k^2}{2m} - \text{Re} \Sigma_k(\omega) \big]^2 + \big[\text{Im} \Sigma_k(\omega) \big]^2} \, .
\label{eq:asf_self}
\end{align}
The real part of the self-energy, $\text{Re} \Sigma_k(\omega)$, is often interpreted as an energy-dependent mean-field potential  \cite{Jeukenne1976}. 
Systems with a well-defined quasi-particle structure have spectral functions with strong peaks around the energy for which the denominator of Eq.~(\ref{eq:asf_self}) is smallest. This happens at the so-called quasi-particle energy, which can be accessed from $\text{Re} \Sigma_k(\omega)$ by solving the recurrent equation 
\begin{align}
\varepsilon_k = \frac{k^2}{2m} + \text{Re} \Sigma_k(\varepsilon_k ) \, ,
\label{eq:qppeak}
\end{align}
at each momentum, $k$. This on-shell condition provides a one-to-one correspondence between energy and momenta - a dispersion relation that is often used to characterise relevant single-particle properties. Other many-body approaches, like many-body perturbation theory or the BHF diagrammatic method, make use of such on-shell dispersion relations throughout \cite{Fetter_book}. The SCGF method, in contrast, considers all energy-dependent, offshell components in the self-energy and the spectral function. These effects are relevant to describe the fragmentation of single-particle strength \cite{Dickhoff_book}. 

The imaginary part of the self-energy is usually discussed in terms of quasi-particle lifetimes \cite{Dickhoff_book}. For the retarded self-energy, the imaginary components is negative, and is directly related to the available phase space.  Indeed, for an energy-independent self-energy, the Fourier transform of Eq.~(\ref{eq:asf_self}) to the time domain provides a propagator that decays exponentially in time with a characteristic timescale,
\begin{align}
\tau_k^{-1} = \Gamma_k= -2 \, \text{Im} \Sigma_k(\varepsilon_k) \, .
\label{eq:lifetime}
\end{align}
The energy and momentum dependence of the self-energy can be characterised in some domains by a model-independent analysis  \cite{Dickhoff_book,Jeukenne1976}. According to Luttinger's theorem for normal fermionic systems, at zero temperature the imaginary part of the self-energy should be exactly zero at the Fermi energy and, close to this energy, it should behave quadratically, so that $\text{Im} \Sigma_k(\omega=\mu) \approx a_k (\omega-\mu)^2$ \cite{Luttinger1961}. When temperature sets in, the $\omega=\mu$ component of Im$\Sigma_k$ grows quadratically with $T$ \cite{Abrikosov1965}.

Expressions for the imaginary part of the self-energy can be derived at different orders in many-body perturbation theory \cite{Dickhoff_book,vanLeeuwen_book}. Compared to their real part counterparts, the imaginary components tend to be simpler to evaluate thanks to energy and momentum conservation. In SCGF methods, one generally computes the imaginary part first, whereas the real-part is obtained from a dispersion relation,
\begin{align}
\text{Re} \Sigma_k(\omega) = \Sigma_k^\infty - \mathcal{P} \int \frac{\textrm{d} \omega'}{\pi} \frac{ \text{Im} \Sigma_k(\omega')}{ \omega-\omega'} \, ,
\label{eq:dispersion}
\end{align}
where $\mathcal{P}$ denotes the principal part. 
The energy-independent (but momentum-dependent) component, $\Sigma_k^\infty$, is akin to a Hartree-Fock self-energy, but includes propagator renormalisation effects via a correlated momentum distribution, 
\begin{align}
\Sigma_k^\infty = 
\nu \int  \frac{\text{d}^3 k_1}{(2 \pi)^3} \langle k k_1 | V | k k_1 \rangle_A n_{k_1}
+ \frac{\nu }{2} 
\int  \frac{\text{d}^3 k_1}{(2 \pi)^3} \int  \frac{\text{d}^3 k_2}{(2 \pi)^3} \langle k k_1 k_2 | W | k k_1  k_2 \rangle_A n_{k_1} n_{k_2} \, .
\label{eq:order1}
\end{align}
In this expression, the matrix elements correspond to ket antisymmetrized two-nucleon ($V$) and three-nucleon ($W$) interactions, respectively. The interactions that are used in this work are discussed in more detail in the following Section.

The self-energy operator can be defined diagrammatically and encodes different many-body processes depending on the level of approximation \cite{Dickhoff_book}. At the lowest order, the so-called mean-field or Hartree-Fock approximation, the self-energy is energy-independent and formally given by the same expression as in Eq.~(\ref{eq:order1}), but with internal momentum distributions, $n_k$, computed from on-shell Fermi-Dirac distributions, $n_k=f(\varepsilon_k)$. Second-order self-energies are instead energy-dependent, proportional to $V^2$ and have already a non-trivial imaginary part \cite{Dickhoff_book,Carbone2020}. These account for 2p-1h and 1p-2h excitations within the system. In infinite matter applications, non-perturbative approximations beyond the second order are typically considered. A self-energy arising from a ladder resummation of the in-medium interaction deals effectively with the strong short-range and tensor components of the NN force, and leads to stable numerical results even for hard core interactions - for details see Refs.~\cite{Frick2003,RiosPhD}. 

When a SCGF calculation converges for a given set of external parameters, one typically stores the complex self-energy, $\Sigma_k(\omega)$. All other microscopic properties, including spectral functions, can be derived from it starting from Eq.~(\ref{eq:asf_self}). The energy per particle of the system is then accessed by an energy-weighted integral of the spectral functions, the so-called Galitskii-Migdal-Koltun sum-rule \cite{Dickhoff_book},
\begin{align}
\frac{E}{A} = 
\frac{\nu}{\rho} \int  \frac{\text{d}^3 k_1}{(2 \pi)^3}  \int  \frac{\text{d} \omega}{ 2 \pi}  \frac{1}{2} \left[ \frac{k^2}{2m} + \omega \right] \mathcal{A}_k(\omega) f(\omega) 
- \frac{1}{2} \langle W \rangle \, .
\end{align}
Here, we also incorporate a correction factor proportional to the expectation value of the 3NF, $\langle W \rangle$. This is computed by using an additional integral over momentum of the second term in Eq.~(\ref{eq:order1}). In other words, we keep the lowest order approximation to the expectation value of the 3NF, in agreement with techniques typically employed in finite nuclei \cite{Cipollone2013}. The formal procedure to go beyond this approximation exists, as discussed in Ref.~\cite{Carbone2013a}, but has not been implemented in infinite systems as of yet.

At finite temperature, the system is described in terms of thermodynamical potentials like the Helmholtz free energy, $F=E-TS$. This requires an explicit calculation for the entropy, $S$, based on Green's functions methods. A proposal to compute $S$ from spectral functions in dense matter was put forward in Refs.~\cite{RiosPhD,Rios2006} based on the formal techniques derived by Pethick and Carneiro \cite{Pethick1973,Carneiro1975}. With access to the entropy, one can then derive all thermodynamical properties, including the pressure, $p=\rho(  F/A - \mu )$. Formally, the SCGF method is known to give rise to thermodynamically consistent results \cite{Dickhoff_book,Kadanoff_book}. For instance, a zero-temperature calculation of the chemical potential, $\mu$, from the Fermi energy of the system, $\varepsilon_{k=k_F}$, or from the density derivative of the free energy density, provide the same result. This formal agreement has also been demonstrated numerically in the past \cite{Rios2006,Soma2009}.

In terms of formalism, there is no inconvenience to extend the SCGF method to asymmetric, multicomponent systems. In fact, the SCGF method has already been used to describe isospin asymmetric matter in the past \cite{Rios2005,Konrad2005}. One can therefore characterise correlations and fragmentation in isospin-imbalanced systems \cite{Rios2009,Rios2014}. In asymmetric matter, all quantities become functions of isospin - eg the spectral functions for neutrons $\mathcal{A}^n_k(\omega)$ and protons  $\mathcal{A}^p_k(\omega)$ are different, as are the self-energies and all other one-body properties. In addition to the isospin splitting, one must also consider changes in the effective interactions, so that neutron-neutron ($nn$), neutron-proton ($np$), proton-neutron ($pn$) and proton-proton ($pp$) in-medium T-matrices are explicitly different. Extensions of the normal ordering procedure to compute effective one and two-body forces from 3NF matrix elements are also necessary~\cite{Carbone2014}, but relatively straightforward~\cite{Wellenhofer2015}. The results for asymmetric matter presented here within the SCGF formalism  include explicitly the effect of isospin asymmetry in the normal-ordering of 3NFs and in the extrapolation procedure from finite to zero temperature.

\subsection{Interactions}

For consistency in the presentation, all the results that follow have been generated with the same set of interactions. I will consider 3 sets of $\chi$EFT-inspired forces. At the NN level, these 3 interactions have been introduced in Ref.~\cite{Coraggio2013}: they have been fitted to the same NN scattering data and incorporate different cutoffs in the relative momentum: $\Lambda=414$ MeV, $450$ MeV and $500$ MeV. These forces also differ slightly in the shape (particularly the sharpness) of the regulator in relative momentum. 3NFs arise naturally in the $\chi$EFT formalism \cite{Epelbaum2009,Machleidt2011}. Reference~\cite{Coraggio2014} introduced a set of N2LO 3NF forces based on the previous 3 NN interactions. The associated 3NF low-energy constants $c_D$ and $c_E$ were fitted to the binding energies of $A=3$ nuclei and to the $^3$H-$^3$He Gamow-Teller matrix elements \cite{Coraggio2014}. Calculations are performed for two-body matrix elements with partial waves up to $J=9$. 

These 3NFs are included in the SCGF simulations by building effective one- and two-body, density-dependent forces using a procedure that is akin to normal ordering \cite{Holt2010,Holt2020}, but includes the renormalisation of strength in $n_k$. This method rooted on the extension of the SCGF formalism to multi-body forces \cite{Carbone2013a}, and has been implemented in symmetric and neutron matter simulations in Refs.~\cite{Carbone2013,Carbone2014}. We resort to some numerical approximations in this averaging procedure. First, the normal-ordering procedure ignores the center-of-mass dependence of the 3NF, which is set to $P=0$. Second, matrix elements off the diagonal in relative momentum, $q \neq q'$, are extrapolated from diagonal matrix elements using the prescription $q \to \frac{1}{2}( q^2 + q'^2 )$. This procedure also involves a 3NF regulator,  which is chosen to be non-local, $f(p,q) = \exp \left[-\left( \frac{q^2 + p^2/3}{\Lambda^2} \right)^{n} \right]$, and based on the two-body relative momentum $q$ and the internal, integrated single-particle momentum, $p$. The exponent $n$ is chosen in accordance to the regulator in the 2NF \cite{Coraggio2014}. Within the $P=0$ approximation, the off-shell $q$ prescription provides results of very similar quality to many-body perturbation theory calculations that incorporate fully-fledged 3NFs from the outset \cite{Drischler2016}. 

\subsection{Extrapolation procedure to zero temperature }

Finite-temperature calculations are useful on their own, but benchmark comparisons with other methods are typically performed in the zero-temperature limit \cite{Baldo2012,Piarulli2020}. A zero-temperature extrapolation procedure for SCGF simulations was developed in the context of beyond-BCS pairing \cite{Ding2016,Rios2017b}. This procedure is now available generically and can be used to provide zero-temperature data for symmetric and asymmetric systems. 
For a given fixed density, the extrapolation requires as input a few (typically of order $4-10$) finite-temperature simulations. One typically performs simulations from high temperatures of order $20$ MeV, and subsequently dials down the temperature to values that are closer to zero. Too close to zero, though, the simulations will not converge as one enters into the superfluid regime - so an intermediate range must be found. It is also important to note that the dependence in temperature typically scales with the dimensionless degeneracy parameter, $T/\varepsilon_F$, where $\varepsilon_F$ is the Fermi energy. Because $\varepsilon_F$ typically increases with density, the expansion is closer to the zero-temperature result (and therefore more accurate) at higher than at lower densities.

\begin{figure}
\begin{center}
\includegraphics[width=1\linewidth]{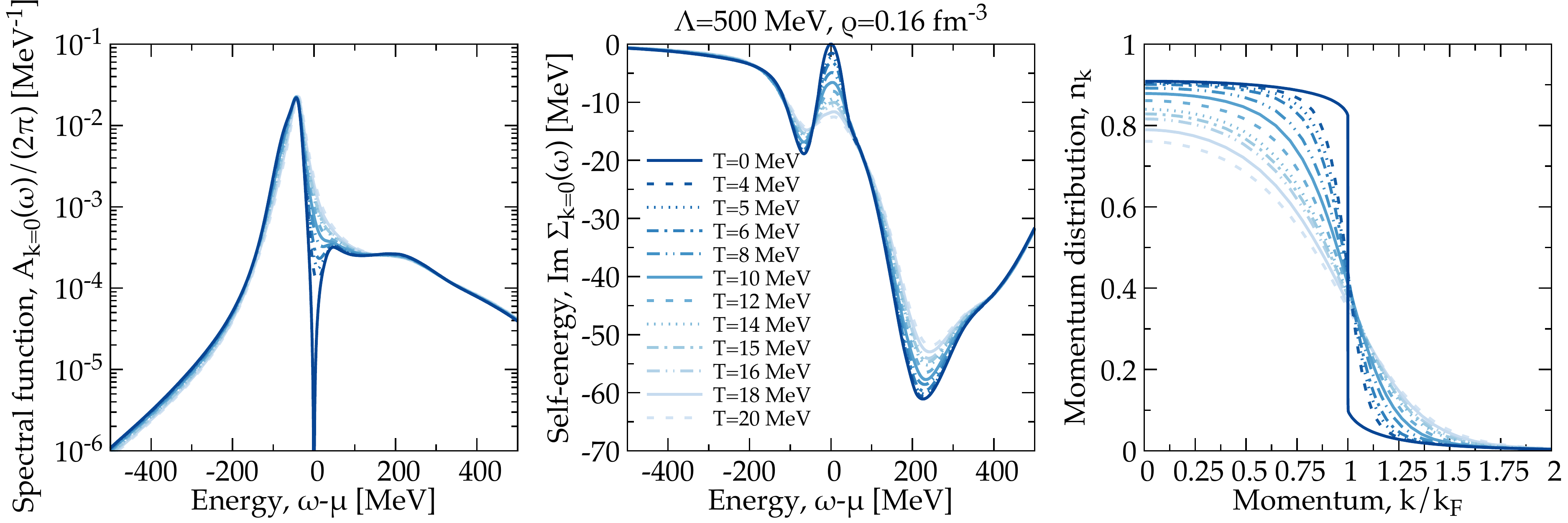}
\end{center}
\caption{
\label{fig:extrapolation} 
Left panel: energy dependence of the zero-momentum spectral function $\mathcal{A}_{k=0}(\omega)$ at different temperatures. Central panel: the same for Im $\Sigma_{k=0}(\omega)$. Right panel: momentum distribution as a function of $k/k_F$. All results have been obtained in symmetric matter, at a density of $\rho=0.16$ fm$^{-3}$for the $\Lambda=500$ MeV N3LO NN plus N2LO 3NFs forces.}
\end{figure}

The extrapolation from finite temperature to zero temperature works in 3 independent steps. In a first step, the self-energies from different temperatures are remeshed to the same values of momentum and energy. In the second step, the temperature dependence of the real and imaginary parts of the self-energy (for each value of $k$ and $\omega$) are fitted to a polynomial function in temperature. The third step is to use not one, but different polynomial fits to extrapolate to $T \to 0$. A series of tests are run on these different extrapolations, and the optimal fit (in terms of thermodynamical consistency) is selected. Additional details of this extrapolation procedure are discussed in Ref.~\cite{Ding2016}. 

I provide an example of this extrapolation procedure in Fig.~\ref{fig:extrapolation}. This shows the temperature dependence of 3 properties that are representative of the system's behavior as a function of temperature at a fixed density of $\rho=0.16$ fm$^{-3}$. Calculations have been performed with the N3LO $\Lambda=500$ MeV NN force, supplemented with 3NFs at the N2LO level. The averaging procedure for these 3NFs consistently takes into account the density and temperature dependence of the internal propagators. Qualitatively similar results are obtained for the $\Lambda=450$ and $414$ MeV interactions. 

The energy dependence of the spectral function $\mathcal{A}_{k=0}(\omega)$  and the imaginary part of the self-energy $\Sigma_{k=0}(\omega)$ are displayed in the left and central panels, respectively. I focus on the $k=0$ value for simplicity, but the calculation and extrapolation procedure are performed for all available momenta. The effect of temperature in the spectral function is concentrated in the region near $\omega=\mu$. The quasi-particle peak and the positive and negative high-energy tails are insensitive to thermal effects. For the self-energy, there is also a relatively strong temperature dependence near $\omega=\mu$, which is known to scale with  $T^2$ \cite{Abrikosov1965}. The phase-space dominated peaks in the particle and hole domains at energies $\approx 200$ MeV away from $\mu$ are also sensitive to the temperature. The extrapolated $T=0$ result is able to capture the temperature dependence of these quantities, while providing consistent results. 

The right panel of Fig.~\ref{fig:extrapolation} shows the associated momentum distribution, $n_k$, obtained in the same conditions. Thermal effects provide additional depletion at low momenta and populate moderately the vicinity of the Fermi surface. In fact, temperature modulates substantially the momentum distribution around $k=k_F$ even for relatively low values of $T \approx 4-5$ MeV. The extrapolation procedure, however, captures the  energy- and momentum-information sufficiently well as to produce a sharp transition at the Fermi surface, as expected from Luttinger's theorem \cite{Abrikosov1965}. The size of this discontinuity is also in agreement with the renormalisation parameter associated to the self-energy itself. This extrapolation procedure has been implemented not only in the symmetric and pure neutron matter cases \cite{Rios2017b}, but also in asymmetric infinite matter, as I shall discuss in the following Sections.

\section{Microscopic properties}
\label{sec:micro}

In the following, I provide an overview of the microscopic properties predicted by the SCGF approach for symmetric, asymmetric and neutron matter. The aim of this section is to show the broad scope of information that can be generated in SCGF simulations. I focus for simplicity on a single value of the density, $\rho=0.16$ fm$^{-3}$, which is representative of both nuclear systems and of the core of neutron stars. Note however that the method is customarily used at other densities. The majority of the results are presented in their zero-temperature extrapolated form, although the temperature dependence of most quantities is available and has been used for extrapolation purposes. I leave a full analysis of the density, isospin asymmetry and temperature dependence of these results for future studies. 

\subsection{Spectral functions}

\begin{figure}
\begin{center}
\includegraphics[width=1\linewidth]{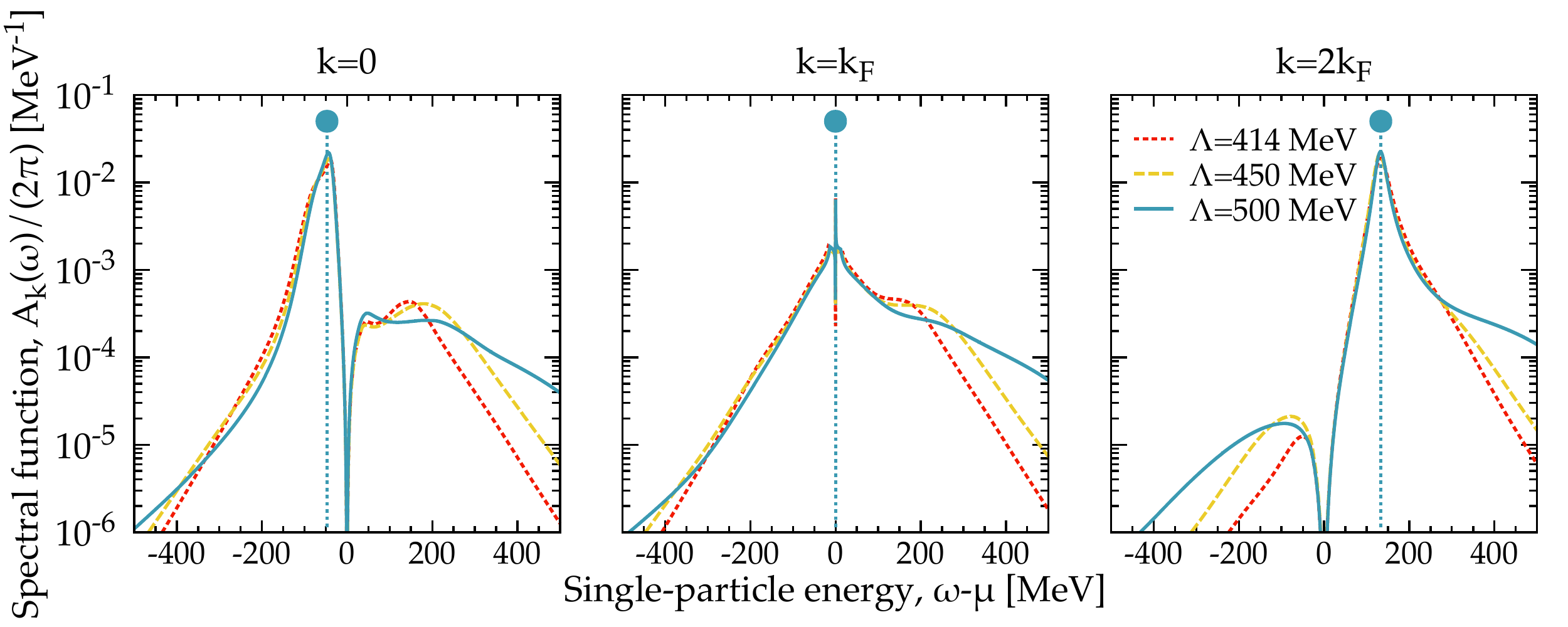}
\end{center}
\caption{
\label{fig:spectral_function} 
Spectral function $\mathcal{A}_k(\omega)$ as a function of energy for three different momenta: $k=0$ (left), $k=k_F$ (central) and $k=2k_F$ (right panel). SCGF calculations are performed at a density $\rho=0.16$ fm$^{-3}$ and have been extrapolated to zero temperature. Results for N3LO NN plus N2LO 3NFs have been performed at three different cutoffs. The vertical line and solid dot indicate the position of the quasi-particle peak, $\omega=\varepsilon_k$, at each one of these momenta. The heigh of this peak has been chosen arbitrarily.}
\end{figure}

Figure~\ref{fig:spectral_function} provides an example of the structure of the zero-temperature spectral function in extended nuclear systems. The three panels illustrate the energy dependence of the spectral function at three different momenta: $k=0$, below the Fermi surface; $k=k_F$, at the Fermi surface; and $k=2k_F$, above the Fermi surface. Results are shown for 3 interactions: $\Lambda=414$ MeV (red dotted lines);  $\Lambda=450$ MeV (yellow dashed lines); and $\Lambda=500$ MeV (blue solid lines). 

The general features of the spectral strength in infinite systems are well understood \cite{Ramos1991,Polls1995,Frick2003,Rios2009b}. At all momenta other than $k_F$, the spectral strength has a prominent quasi-particle peak. In infinite nuclear matter, the peak is very close to the quasi-particle energy $\omega=\varepsilon_k$ obtained from Eq.~(\ref{eq:qppeak}). The quasi-particle energy for the $\Lambda=500$ MeV interaction is displayed in the 3 panels of Fig.~\ref{fig:spectral_function} with vertical lines. The solid dot is there for illustrative purposes, and its height is chosen arbitrarily. The quasi-particle peak is well-defined in infinite nuclear systems, in the sense that the peak of $A_k(\omega)$ and $\omega=\varepsilon_k$ agree relatively well. At this density, the quasi-particle peak is independent of the interaction under consideration. In fact, the cut-off dependence in the quasi-particle peak is not resolvable in the scale if this graph. I note, however, that while the peaks are well-defined, they are also relatively broad. Typical widths at $k=0$ and $k=2k_F$ are of the order of $\Gamma_k \approx 40$ MeV.

Spectral functions in infinite matter display slowly decaying high (positive and negative) energy tails. These are particularly prominent at the Fermi surface, $k=k_F$, where the zero-temperature spectral function is a combination of a broad energy-dependent background and a zero-width $\delta$ function centred at $\omega=\varepsilon_{k_F}$. 
The results of Fig.~\ref{fig:spectral_function} clearly show a substantial cut-off dependence in these high-energy components. This is relatively unsurprising, since these tails are a direct consequence of the short-range core in the NN force \cite{Dickhoff2004}. As expected, the interaction with the largest momentum cut-off ($\Lambda=500$ MeV), and hence the largest resolution in real space, provides the largest high-energy components. These includes a shoulder at positive energies in the region $100-500$ MeV which is characteristic of a  relatively strong short-range core. In contrast, the lower cut-off interactions provide much faster (exponential) decaying spectral functions beyond the quasi-particle peak. For positive energies, they provide very little single-particle strength beyond $500$ MeV. The $\Lambda=414$ MeV interaction has faster decaying tails than the $450$ MeV force, which bodes well with intuitions. Note that this sharply decaying spectral strength is at odds with calculations employing traditional phenomenological NN forces, which typically populate these offshell regions with higher probabilities \cite{Frick2003,Rios2017}. Off-shell high-energy components are observed in electron-scattering experiments \cite{Rohe2004} and may provide a way to quantify the short-range component of  NN forces   \cite{Schmidt2020}.

The results presented so far concerned symmetric nuclear matter, where by construction one assumes the same fraction of neutrons and protons. Isospin asymmetric systems, however, are interesting on their own. The deep interior of heavy stable and unstable isotopes is expected to be a neutron-rich environment, characterised by an isospin asymmetry parameter, 
\begin{align}
\eta=\frac{\rho_n-\rho_p}{\rho_n+\rho_p} \, ,
\end{align}
of order $\eta \approx 0.2$ \cite{TypelBrown2001}. In contrast to this relative mild isospin asymmetry, the core of neutron stars is expected to be in extremely asymmetric conditions with  $\eta \approx 0.8$ \cite{Haensel}. Finite-temperature SCGF simulations including explicitly isospin asymmetry have been available for over a decade \cite{Rios2005}. In these simulations, $\eta$ is an external, tuneable parameter. SCGF calculations then give isospin-dependent self-energies, spectral functions and quasi-particle properties, as well as the corresponding bulk properties. These calculations have provided insight into the isospin-asymmetry dependence of the fragmentation of single-particle strength \cite{Rios2009,Rios2014}, 
but were somewhat limited by two factors. First, finite temperature effects have a larger importance in the minority component as asymmetry is turned in, because the corresponding degeneracy parameter $\frac{T}{\varepsilon_F^\tau}$ decreases. Second, the calculations were performed with two-body interactions only and thus missed the effect of 3NFs. 
None of these factors is particularly critical. Temperature effects can be controlled over, and the effect of 3NFs on the spectroscopic strength is small below saturation density for both symmetric and neutron matter \cite{Carbone2013,Carbone2014}.

\begin{figure}
\begin{center}
\includegraphics[width=0.8\linewidth]{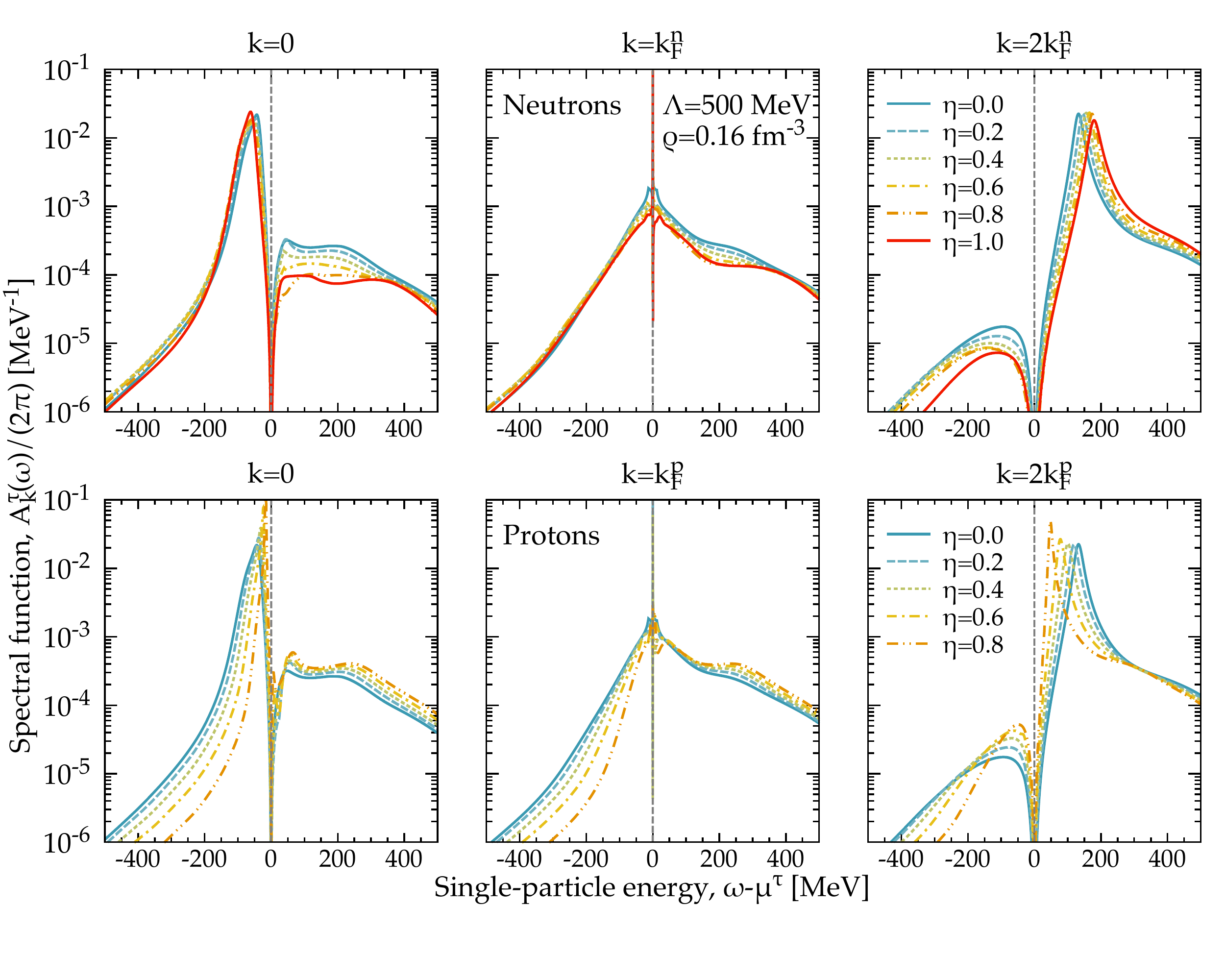}
\end{center}
\caption{
\label{fig:sf_asy} 
Neutron (top panels) and proton (bottom panels) spectral functions $\mathcal{A}^\tau_k(\omega)$ as a function of energy for three different momenta: $k=0$ (left), $k=k^\tau_F$ (central) and $k=2k^\tau_F$ (right panels). Calculations with the N3LO $\Lambda=500$ MeV and an N2LO 3NF at a density $\rho=0.16$ fm$^{-3}$ have been extrapolated to zero temperature for different isospin asymmetries, $\eta$. The vertical dashed line is a guide to separate hole ($\omega < \mu^\tau$) and particle ($\omega > \mu^\tau$) components.}
\end{figure}

Having said that, recent technical developments have allowed us to overcome these two shortcomings recently. Figure~\ref{fig:sf_asy} shows the spectral functions of neutrons (top panels) and protons (bottom panels) as a function of energy for different values of the isospin asymmetry, $\eta$, at a fixed density of $\rho=0.16$ fm$^{-3}$. Results are presented for the $\Lambda=500$ MeV N3LO potential supplemented with 3NFs \footnote{I note here for completeness that the $\eta=0$ and $\eta=1$ calculations are performed with independent codes that do not include isospin asymmetry explicitly, and hence provide a good numerical test of the arbitrary isospin simulations.} The results for the isospin dependence of these spectral functions are qualitatively similar to previously obtained results which lacked 3NFs. The evolution of the neutron spectral function from symmetric ($\eta=0$) to neutron ($\eta=1$) matter is relatively mild. Quasi-particle peaks with respect to the corresponding chemical potential $\mu^n$ become slightly more (attractive) repulsive for the (hole) particle components. The high energy tails are relatively insensitive to isospin, and are mostly affected for the negative energy components at momenta above the Fermi surface. All in all, this bodes well with the generic picture of a neutron majority component evolving from a symmetric situation, where tensor correlations provide larger correlations, to neutron matter, where the short-range component is responsible for fragmentation. 

In contrast, the proton spectral function (bottom panels) is much more affected by isospin asymmetry. In asymmetric conditions, the relative importance of $np$ pairs increases for the minority proton component. Since $np$ interactions are mediated by the tensor component of the NN force, one naively expects enhanced correlation effects for protons in neutron-rich matter. Simultaneously, in neutron-rich conditions the overall proton density decreases, so the proton component should somehow become ``less interacting". SCGF simulations allow us to provide quantitative predictions to distinguish between these scenarios. 

As asymmetry increases, the single-particle peak for protons becomes more repulsive with respect to $\mu^p$. Importantly, the width of the quasi-particle peak for components below the Fermi surface (eg $k=0$ panel) also decreases substantially. As a consequence, positive high-energy tails increase with isospin. Interestingly, the central panel, focused at $k=k_F^p$, shows a background component that behaves differently for the addition and removal energy components. Whereas for $\omega < \mu^p$, the background of the spectral function decreases with isospin asymmetry, at $\omega > \mu^p$, it increases. Above the Fermi surface ($k>k_F^p$), the quasi-particle peak is more compressed with asymmetry, both in terms of a lower peak and a significantly decreased width. The limit case of a proton impurity in an $\eta=1$ system cannot be tackled with the present set of SCGF technology. 

While a full analysis is beyond the scope of this initial presentation of results, the isospin dependence shown here is relevant for both nuclear experiments and astrophysics. 
Hadronic and electron two-nucleon knock-out reactions in from the early 2000s up to present have shown that $np$ pairs, rather than $pp$ or $nn$ pairs, are more likely to be knocked out off isospin symmetric targets  \cite{Tang2003,Piasetzky2006}. Further electron scattering results in isospin asymmetric nuclei have provided further evidence for this neutron-proton dominance \cite{Hen2014}. It is likely that the tensor component preferentially acts in relatively high-momentum components and provides the physical mechanism behind these strong isophobic correlations \cite{Rios2009,Sargsian2014}. Extending the $np$ dominance picture to isospin imbalanced systems would naively indicate that protons should be more correlated than neutrons in neutron-rich systems although, of course, this depends sensitively on the definition of a correlation measure \cite{Duer2018,Ryckebusch2019,Paschalis2020}. In the astrophysical context, the strong modification of the proton spectral function in isospin asymmetric matter predicted by SCGF simulations has been explicitly explored at the level of the symmetry energy \cite{Carbone2012,Hen2015,Cai2016}. The fragmentation of strength in neutron matter has also been used to provide a description of the pairing properties in dense matter \cite{Ding2016,Rios2017b}.

\subsection{Self-energies}
\label{sec:imself}

\begin{figure}
\begin{center}
\includegraphics[width=0.8\linewidth]{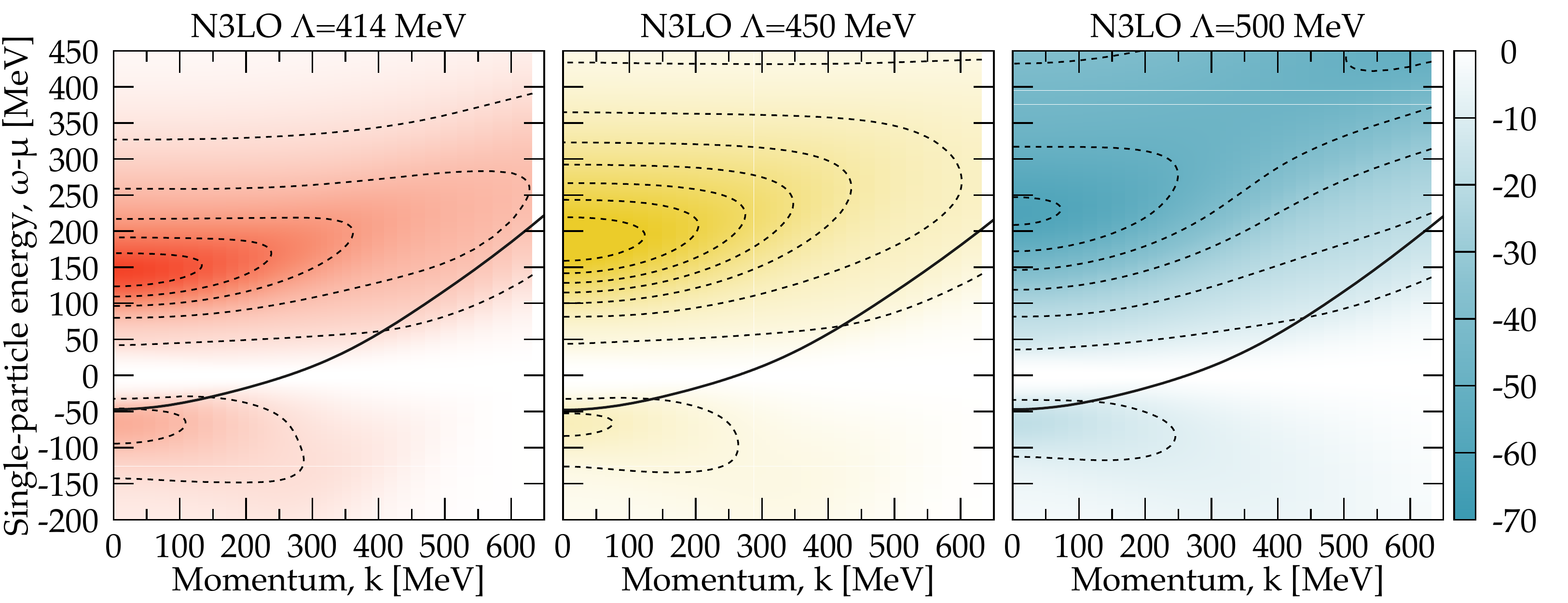}
\end{center}
\caption{
\label{fig:imself} 
Density plot of the imaginary part of the self-energy, $\text{Im} \Sigma_k(\omega)$, as a function of momentum and energy. Dashed lines represent contours at constant values spaced by $10$ MeV. Extrapolated zero-temperature results for symmetric nuclear matter at a density $\rho=0.16$ fm$^{-3}$ are shown for the N3LO NN and N2LO 3NF combinations at cutoffs $\Lambda=414$ MeV (right), $450$ MeV (central) and $\Lambda=500$ MeV (right panel). The solid line represents the quasi-particle peak, $\omega=\varepsilon_k$. 
}
\end{figure}

Within the SCGF formalism, the spectral strength is directly related to the self-energy. The imaginary part of the retarded self-energy encodes information on both the phase space and the NN interaction, and can even be used to diagnose interactions by means of sum rules \cite{Rios2017}. I present a density plot of the zero-temperature extrapolated imaginary part of the self-energy at $\rho=0.16$ fm$^{-3}$ in Fig.~\ref{fig:imself}. Left, central and right panels correspond to chiral interactions with cutoffs $\Lambda=414, 450$ and $500$ MeV including, in all cases, 3NFs. As expected, Im$\Sigma_k$ is zero at $\omega=\mu$ throughout all momenta. For energies $\omega < \mu$, this function has a  hole component with support below $k \approx k_F=263$ MeV.  The size and shape of this pocket is relatively interaction-independent. In contrast, the positive energy ($\omega>\mu$) components of Im$\Sigma_k$ show a clearer cut-off dependence. The deep pocket around $k=0$ in this component is sensitive to the underlying NN force. Typical phenomenological interactions have deep pockets, and these extend to very high positive energies (into the GeV domain). For this set of chiral interactions, the depth of the pocket is relatively insensitive to the cutoff and of order $\approx 50-70$ MeV. The tails of Im$\Sigma_{k=0}$ die off relatively quickly with energy. In fact, the $\Lambda=414$ and $450$ MeV self-energies are almost zero beyond an energy of about $500$ MeV. The cutoff also has a clear effect in the momentum dependence of Im $\Sigma_k$ in the positive energy regions. Generally speaking, higher cutoff forces allow the imaginary part of the self-energy to extend to higher momenta. 

The solid lines in Fig.~\ref{fig:imself} show the trajectory of the quasi-particle peak in the energy-momentum plane. Quasi-particle lifetimes, $\tau_k$ in Eq.~(\ref{eq:lifetime}), are computed along these trajectories. The plot indicates that the lifetimes are relatively insensitive to the differences in self-energies. In other words, largely different offshell self-energies can give rise to relatively similar onshell properties. Onshell components will be discussed in detail in Subsection~\ref{sec:onshell}.

\begin{figure}
\begin{center}
\includegraphics[width=\linewidth]{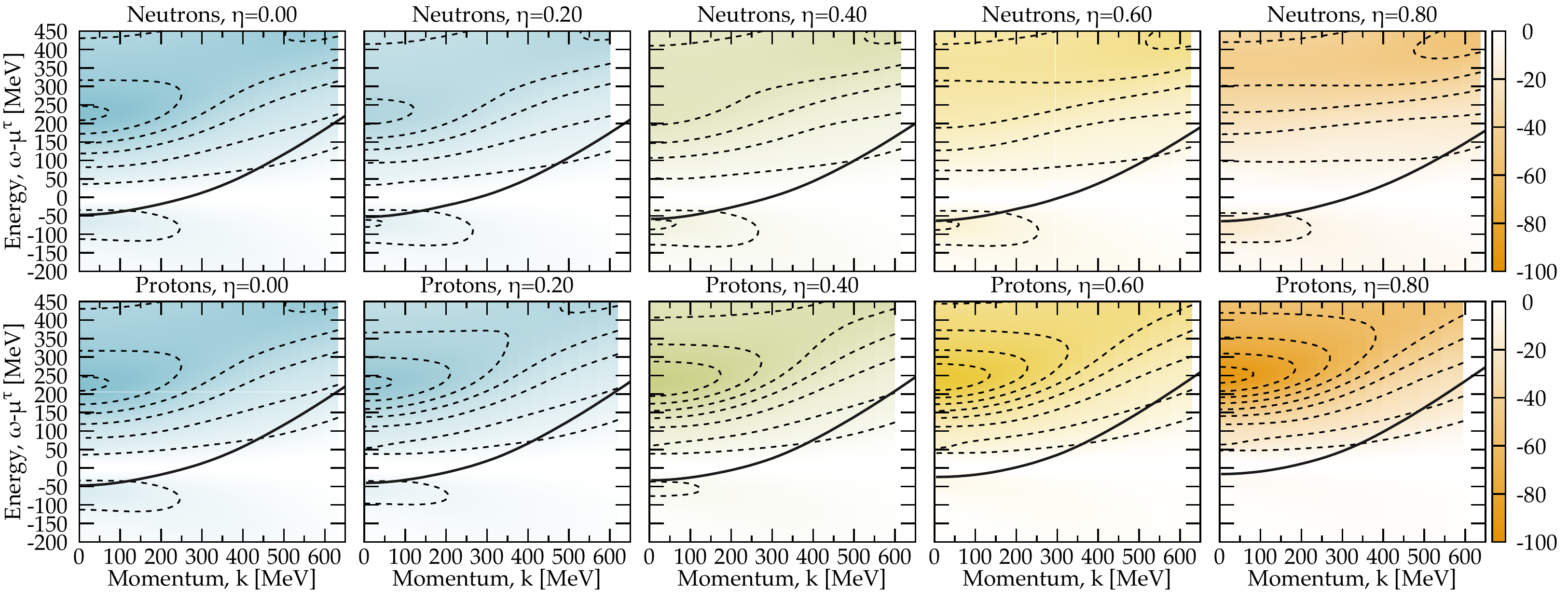}
\end{center}
\caption{
\label{fig:imselfasy} 
The same as Fig.~\ref{fig:imself} but in isospin asymmetric matter for the $\Lambda=500$ MeV interaction. Top panels focus on neutrons and bottom panels, on protons. The panels from left to right correspond to different isospin asymmetries, from $\eta=0$ (symmetric matter, leftmost panel) to the neutron-rich domain with $\eta=0.8$ (rightmost panel). 
}
\end{figure}

The isospin asymmetry dependence of Im$\Sigma^\tau_k(\omega)$ is explored in Fig.~\ref{fig:imselfasy}. Results are shown for the $\Lambda=500$ interaction only, at a fixed density of $\rho=0.16$ fm$^{-3}$, for a variety of isospin asymmetries. The leftmost panel corresponds to the symmetric system ($\eta=0$), and is the same as the right panel of Fig.~\ref{fig:imself}. The top panels indicate that the effect of asymmetry in the neutron self-energy is relatively mild. The pocket at negative energies, $\omega < \mu^n$, associated to neutron holes, is relatively unchanged as $\eta$ increases. In contrast, the positive energy (particle) components evolve with isospin in a substantially different way. The pocket that in the symmetric situation decayed relatively quickly with momentum, becomes an almost momentum-independent structure in the most asymmetric case explored here. 

In contrast, proton self-energies show a strong dependence in asymmetry. As the proton fraction decreases, the negative-energy (hole) components of Im$\Sigma_k^p$ disappear steadily. This reflects the decrease in density, and hence of available phase space, of protons. The corresponding low-momentum ($k<k_F^p$) quasi-particle lifetime is small, which in turn gives rise to an increasingly narrow spectral function. This is clearly in agreement with the results in the bottom left panel of Fig.~\ref{fig:sf_asy}. On the positive energy side of the self-energy, $\omega > \mu^p$, the deep pocket at low momenta becomes noticeably deeper as $\eta$ increases but, unlike the neutron component, the momentum dependence remains relatively unchanged. The increase in depth of Im$\Sigma_k$ can be interpreted in terms of an increase in the interaction strength, as expected in the asymmetric case with an increase in the number of $np$ pairs.

In other words, the imaginary component of the self-energy for protons in asymmetric matter shows the competition of the two effects discussed earlier in two different energy regimes. On the one hand, proton holes, associated to the negative energy components, are dominated by phase-space and their self-energy decreases steadily as the proton density decreases. On the other hand, proton particles at positive energies show an increase in interaction strength as expected in a more imbalanced more strongly interacting system. 
The real part of the self-energy is not shown here for brevity. I note, however, that this is connected to Im$\Sigma_k^\tau$  by the dispersion relation of Eq.~(\ref{eq:dispersion}). One would then obtain equivalent conclusions based on a dispersion relation analysis. 

\subsection{Momentum distributions}

\begin{figure}
\begin{center}
\includegraphics[width=0.7\linewidth]{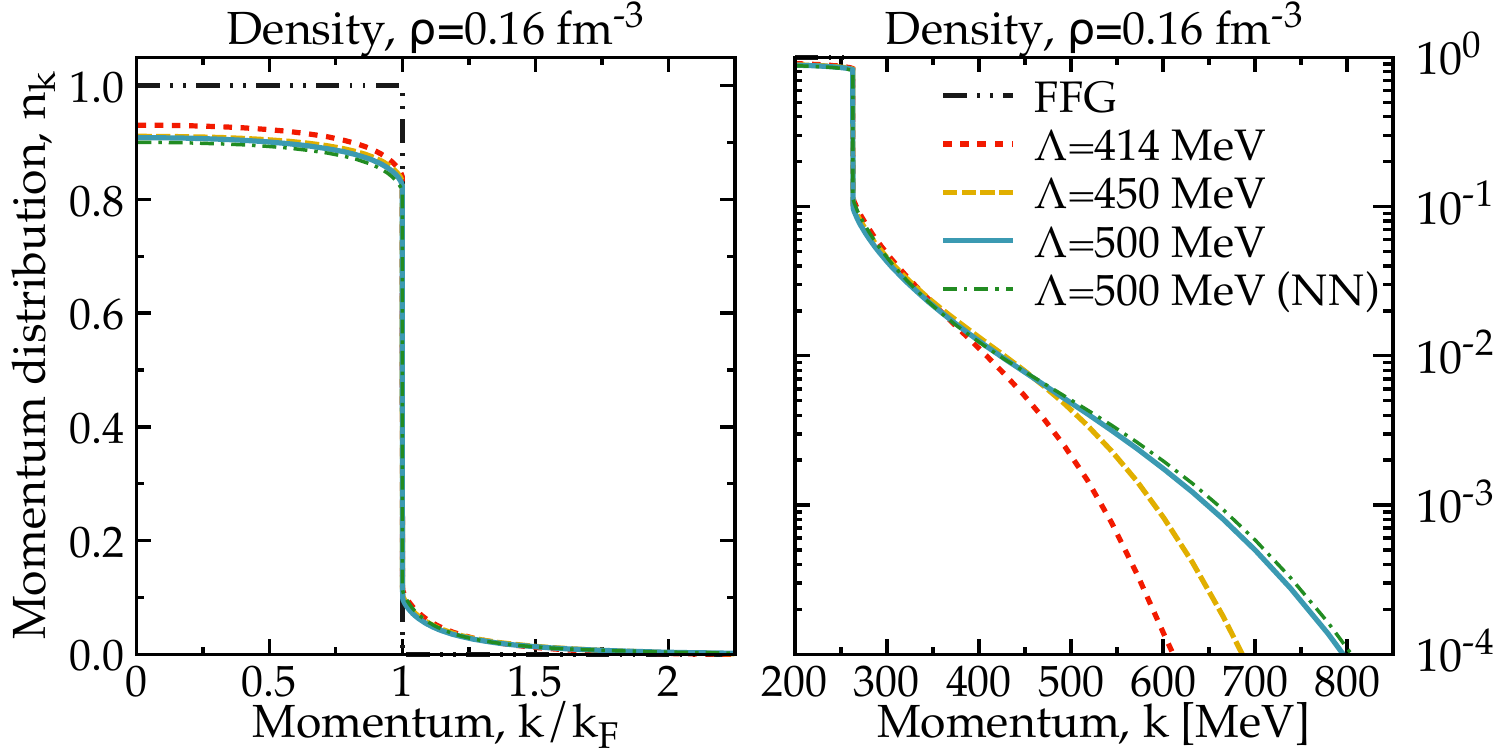}
\end{center}
\caption{
\label{fig:momdisym} 
Extrapolated zero-temperature momentum distribution  obtained with different NN
interactions for symmetric nuclear matter at $\rho=0.16$ fm$^{-3}$.  The left panel shows the distribution in a linear scale as a function of $k/k_F$. The right panel shows the same results in a logarithmic scale and as a function of momentum, $k$, in units of MeV. The black dashed-double-dotted line corresponds to the free Fermi gas (FFG). The green dashed-dotted line is obtained from a calculation with the $\Lambda=500$ MeV interaction without 3NFs. Other lines follow the convention of Fig.~\ref{fig:spectral_function}.
}
\end{figure}

A useful characterisation of beyond mean-field correlations is obtained with the momentum distribution, Eq.~(\ref{eq:momdis}). For a non-interacting Fermi gas or within the lowest-order Hartree-Fock approximation at zero temperature, $n_k=1$ ($0$) for momenta below (above) the Fermi surface, $k_F$. This behaviour is illustrated by the  double-dotted-dashed line in the left panel of Fig.~\ref{fig:momdisym}. In contrast, the momentum distribution of correlated infinite nuclear matter shows a characteristic depletion below the Fermi surface. This depletion is of order $7-12 \%$, with small quantitative differences between interactions. For the 3 chiral interactions under consideration, the cutoff dependence below $k_F$ is very small and well within $\approx 1\%$. 

In turn, a depletion of momentum eigenstates below $k_F$ means that $n_k$ above the Fermi surface must be populated. The logarithmic scale in the right panel of Fig.~\ref{fig:momdisym} illustrates the effects of such high-momentum components. These components are noticeably different depending on the cutoff. Smaller cutoff interactions give rise to momentum distributions that decay faster, and start decaying at lower momenta, than larger cutoff forces. 
To illustrate the relatively small effect that 3NF have on correlations, I show in Fig.~\ref{fig:momdisym} the momentum distribution obtained for a SCGF calculation without 3NFs (dashed-dotted line). Remarkably, the difference between the calculation with and without 3NFs is substantially smaller than the cutoff dependence of the interaction. This bodes well with previous observations indicating that the effect of 3NFs on fragmentation properties is quantitatively small, even though they have a large influence on the energetics and thermodynamics of the system \cite{Carbone2013}.

\begin{figure}
\begin{center}
\includegraphics[width=0.7\linewidth]{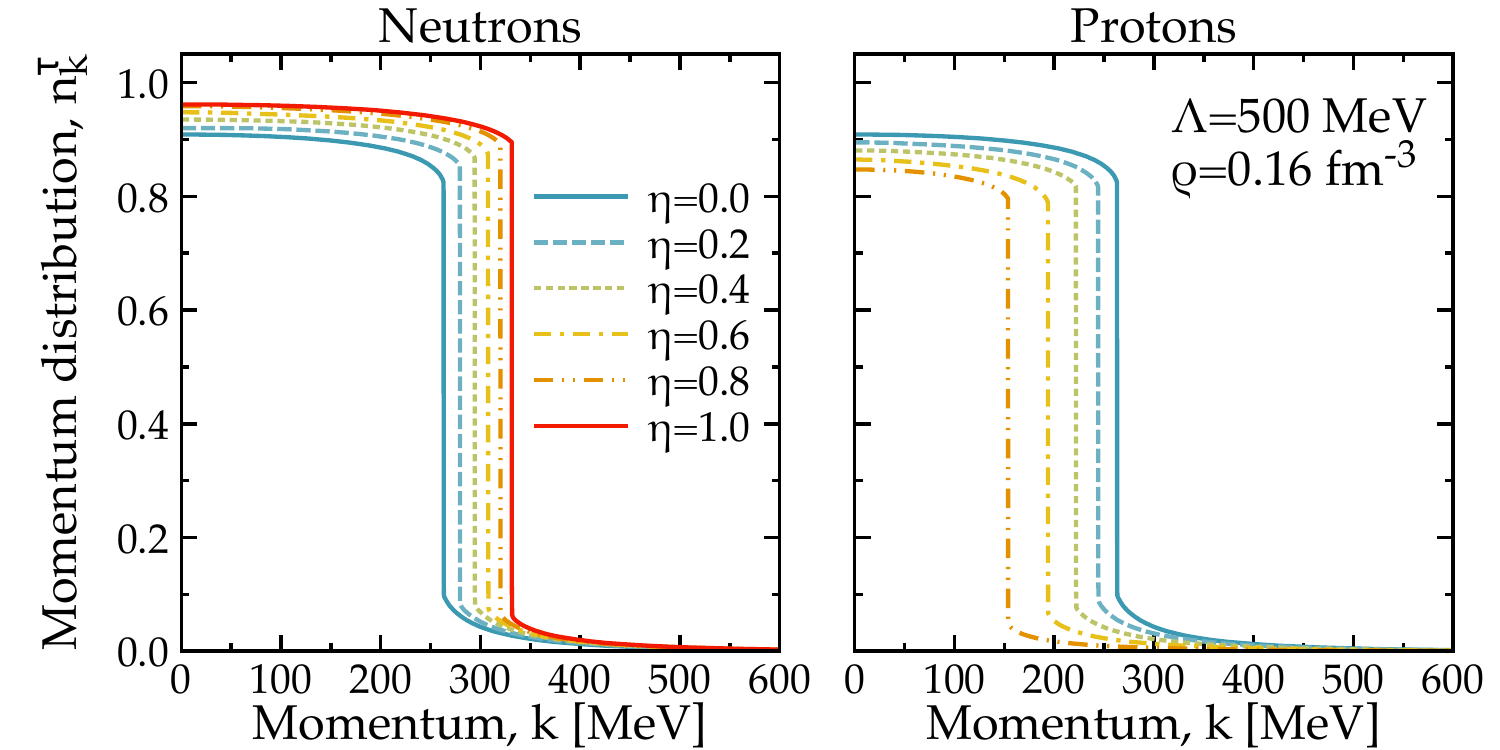}
\end{center}
\caption{
\label{fig:momdisasy}
Left panel: neutron momentum distribution, $n^n_k$, as a function of momentum. Extrapolated zero-temperature calculations at different values of isospin asymmetry, $\eta$, are shown at a fixed density of $\rho=0.16$ fm$^{-3}$. Results have been obtained for the $\Lambda=500$ MeV interaction including 3NFs. Right panel: the same for the proton momentum distribution $n^p_k$.
}
\end{figure}

The evolution of the momentum distribution with isospin also provides an illustrative understanding of the interplay between short-range and tensor correlations and isospin asymmetry. Figure~\ref{fig:momdisasy} shows the momentum distribution as a function of momentum for different isospin asymmetries. These results have been obtained with the $\Lambda=500$ MeV interaction, but qualitatively similar plots are found for the other 2 forces. The left panel shows the results for neutrons. Isospin asymmetry has two main effects on the majority component. On the one hand, the discontinuity in $n_k$ moves to higher momenta, as the Fermi momentum changes from $k_F=263$ MeV in symmetric matter to $k_F=331$ MeV in neutron matter.  This effect would be identical in the FFG. In contrast, the distinct increase of $n_k$ below the Fermi surface is specific to correlated systems. The depletion of strength decreases (or gets closer to $1$) linearly with the increase of asymmetry. In neutron matter, the depletion goes from around $10\%$ in symmetric matter down to a level of around $4-5 \%$ in neutron matter. In fact, this isospin dependence is rather universal and independent of the NN interaction under consideration \cite{Rios2009,Rios2014}. A similar effect is expected for the size of the discontinuity across the Fermi surface, which decreases as isospin asymmetry increases. 

The right panel of  Fig.~\ref{fig:momdisasy} provides insight on the asymmetry evolution of the minority component (protons). Mirror effects are obtained in this case. The discontinuity of $n_k$ occurs at smaller and smaller momenta, as expected from FFG considerations. The depletion below the Fermi surface departs further away from $1$ as isospin asymmetry increases. In other words, by this measure, protons (or minority components) becomes more correlated in a neutron-rich environment. Close to the astrophysical relevant conditions at $\eta \approx 0.8$, protons are depleted by about $15 \%$ for this specific interaction. This depletion goes in hand with a change in high-momentum components for the proton, which become relatively more important as asymmetry increases \cite{Rios2014}. 

The picture obtained by these complete calculations including zero temperature extrapolations and 3NFs is qualitatively and quantitatively very similar to that already discussed in Refs.~\cite{Rios2009,Rios2014}. Overall, the effect of isospin asymmetry on occupation numbers is relatively small. For the characteristic isospin asymmetry in the interior of nuclei ($\eta=0.2$), the change in occupation below the Fermi surface is of the order of only a few percent. Overall, this shallow asymmetry dependence is at odds with some of the past and recent experimental analysis in both electron- and hadronic-induced knock-out reactions \cite{Gade2008,Duer2018,Paschalis2020}. Importantly, however, these calculations at fixed density do not account for the density and isospin-asymmetry dependence within nuclei, which is likely to be relevant in the description of nuclear momentum distributions. 

\subsection{Onshell properties}
\label{sec:onshell}

\begin{figure}
\begin{center}
\includegraphics[width=\linewidth]{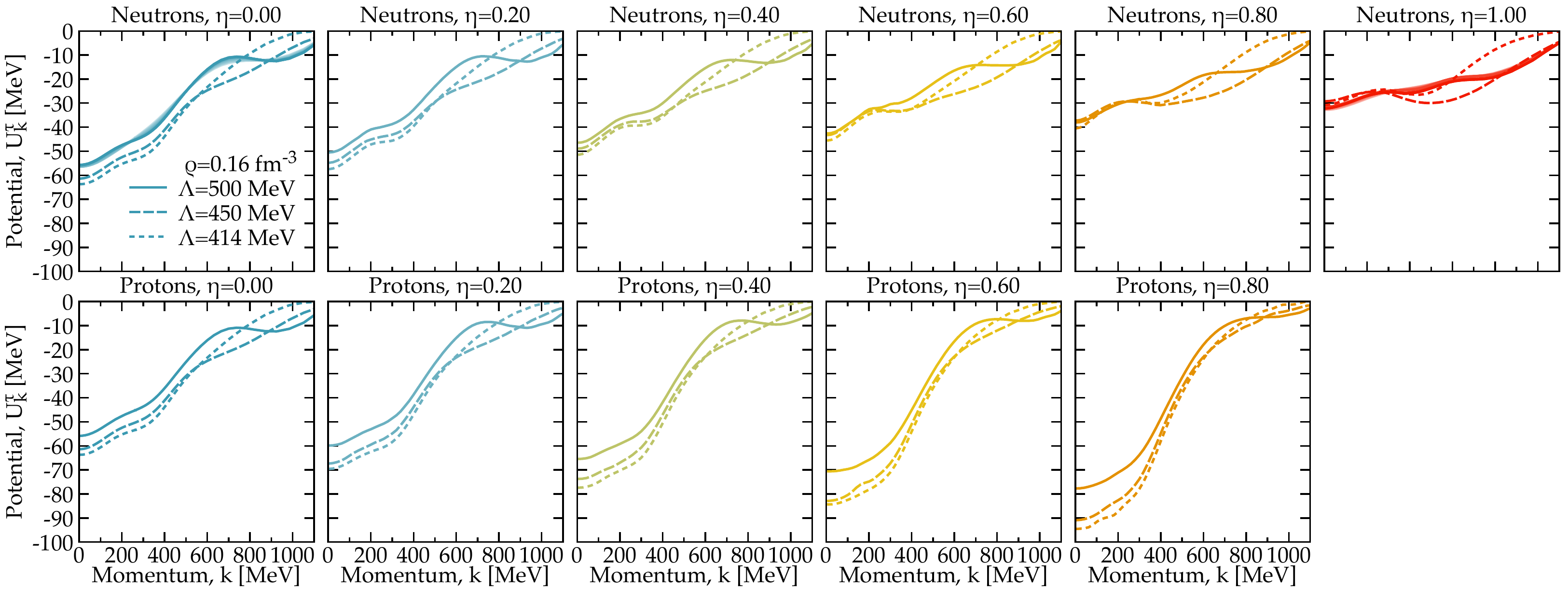}
\end{center}
\caption{
\label{fig:qp_asy} 
Neutron (top panels) and proton (bottom panels) single-particle potentials as a function of momentum for the 3 interactions considered in this work. The panels from left to right correspond to different isospin asymmetries, from $\eta=0$ (symmetric matter, leftmost panel) to neutron matter (rightmost panel). The band for the $\eta=0$ and $\eta=1$ $\Lambda=500$ MeV results displays the temperature dependence from $T=20$ to $T=0$ MeV.
}
\end{figure}

A unique advantage of SCGF simulations is the access to both off-shell, energy-dependent quantities, as well as the corresponding information on the energy shell - at the quasi-particle energy $\omega=\varepsilon_k$. On-shell properties are often used in interpreting the properties of the system and are the inputs to most astrophysical simulations. In particular, quasi-particle potentials in asymmetric matter at relatively low (neutrino-sphere) densities are important for charged-current neutrino opacities \cite{MartinezPinedo2012,Roberts2012}.  The quasi-particle spectrum of Eq.~(\ref{eq:qppeak}) is a key quantity that allows for a characterisation of single-particle excitations in the system. Since the kinetic term in the quasi-particle energy obscures the many-body effects at large momentum, I instead focus on the single-particle potential $U_k^\tau = \text{Re} \Sigma_k^\tau(\varepsilon_k^\tau)$. This is shown as a function of momentum in Fig.~\ref{fig:qp_asy}. Top panels correspond to neutrons, whereas bottom panels correspond to protons. From left to right, the system changes isospin asymmetry from symmetric matter ($\eta=0$) to neutron matter ($\eta=1$) at a constant density of $\rho=0.16$ fm$^{-3}$. I show results for the $\Lambda=500$ (solid), $450$ (dashed) and $414$ MeV  (dotted lines) interactions.

As discussed previously, these results are obtained from a zero-temperature extrapolation of finite temperature SCGF simulations. As an illustration of the importance of thermal effects, I also show in the top left and top right panels the $\Lambda=500$ MeV results for $U^\tau_k$ at temperatures up to $T=20$ MeV. The thermal effects are clearly small in the scale of this figure, particularly in comparison to the cutoff and the isospin dependence of the results. I note in passing that the temperature dependence is non-monotonous. At momenta below the Fermi surface, $U^\tau_k$ becomes more repulsive as temperature increases. Above the Fermi surface and up to about $\approx 500$ MeV, the potential instead becomes more repulsive with temperature. This in agreement with previously reported SCGF results that did not incorporate 3NFs~\cite{Rios2009b}. 

In terms of isospin dependence, the results also bode well with previously reported data in the SCGF approach \cite{Rios2005}, and in other many-body calculations \cite{Zuo2002,Zuo2005,Holt2016}. In symmetric matter, the single-particle potential has a minimum at low momenta close to $U_{k=0} \approx -60$ MeV, and increases steeply with momenta. At $k=0$, the lowest cutoff interaction produces a potential that is about $\approx 5$ MeV more attractive than the highest cutoff force. This cutoff-dependence is relatively constant as a function of momentum up to about $k\approx600$ MeV, where  regulator artefacts start to dominate. Above this value of momentum, the $414$ and $450$ MeV force results go to zero relatively quickly, whereas the potential for the $500$ MeV interaction shows a characteristic shallow minimum around $800$ MeV. 

As asymmetry increases, the neutron single-particle potential becomes more repulsive and shallower in terms of momentum dependence. In contrast, protons experience a larger attraction when they become the minority species. Their single-particle potentials become significantly more attractive and, in the astrophysically relevant region of $\eta \approx 0.8$, they reach values of $80-90$ MeV at $k=0$. The isospin dependence is to a good approximation linear with isospin, as expected on general grounds \cite{Zuo2005}. I note however that the momentum and isospin dependence of these single-particle potentials is significantly different from those predicted by mean-field phenomenological momentum-dependent potentials \cite{Sellahewa2014}.

Interestingly, the cutoff dependence of the results depends on isospin and on the nuclear species. As neutrons become the predominant species and their single-particle potentials become more shallow, the cutoff dependence in the low-momentum single-particle potential below $\approx 400$ MeV decreases. In neutron matter, the differences between the potentials are of the order of $\approx 1$ MeV. In stark contrast, as the isospin asymmetry increases and the proton spectrum becomes more attractive, the differences between interactions with different cutoffs increase and become of the order of more than $10$ MeV. 

These differences can be explained by two interrelated factors. One is of a purely practical nature. $np$ interactions play no role in neutron matter, and hence the intrinsic differences in the NN interaction are less likely to appear as several partial waves channels are suppressed. In other words, if the fits to the $nn$ channels have intrinsically smaller differences between forces than the corresponding $np$ channels, neutron-rich systems will show less cut-off dependence. Another factor that plays an important role is the strength (in the sense of ``perturbativeness") in the different channels. In terms of $\chi$EFT, the small cutoff dependence in neutron matter bodes well with the idea that neutron matter is a more perturbative system than symmetric matter is \cite{Coraggio2013,Hebeler2013,Drischler2014}, precisely because the strong $np$ components are missing. It is important to emphasise here that the strong cutoff dependence of the impurity spectrum may not be identified in the energetics of the system, but will clearly have an impact on the dynamics of asymmetric systems. 

The momentum dependence of $U_k^\tau$ is directly related to the concept of effective mass. SCGF simulations can be used to extract not only the mass associated to the single-particle spectrum, but also to the so-called $\omega-$ and $k-$effective masses. A detailed analysis of these properties within the SCGF approach is ongoing. 

\begin{figure}
\begin{center}
\includegraphics[width=\linewidth]{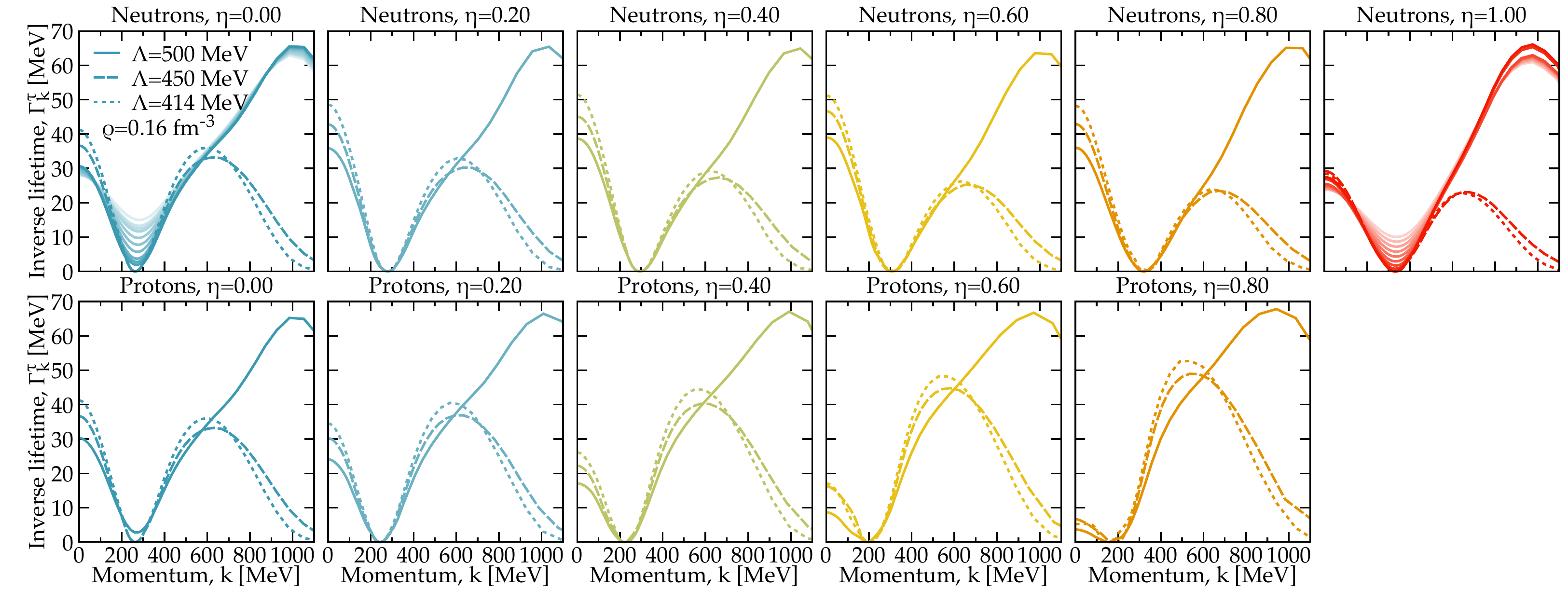}
\end{center}
\caption{
\label{fig:gamma_asy} 
Neutron (top panels) and proton (bottom panels) inverse single-particle lifetimes as a function of momentum. For details, see the caption of Fig.~\ref{fig:qp_asy}
}
\end{figure}

The on-shell imaginary part of the self-energy, Eq.~(\ref{eq:lifetime}), can be interpreted as an inverse quasi-particle lifetime. In a well-defined many-body theory for a normal correlated system at zero temperature, $\Gamma_{k=k_F}$ should be zero, thus signalling an infinite lifetime for quasi-particles at the Fermi surface. The lifetime should be non-zero everywhere and, as discussed in the context of Subsection~\ref{sec:imself}, its magnitude is a combination of phase-space and interaction effects. Figure~\ref{fig:gamma_asy} provides an overview of the momentum dependence of $\Gamma_k^\tau$ for neutrons (top panels) and protons (bottom panels) as a function of isospin asymmetry. As in the previous figure, extrapolated zero-temperature results are shown. 

The isospin and cutoff dependence of the results is very different above and below the Fermi surface. Below the Fermi surface, the SCGF predict a non-zero width. Unlike BHF calculations, intermediate hole states explicitly account for this non-zero width for hole states, $k<k_F$ \cite{Zuo2005}. For neutrons, this hole component is relatively independent of asymmetry. In fact, $\Gamma^n_{k=0}$ increases from values around $30-40$ MeV in symmetric matter to $40-50$ MeV around $\eta=0.6$, only to decrease again to values around $20-30$ MeV in neutron matter. In contrast, for protons the hole component decreases steadily with asymmetry. This reflects the decrease in phase space of protons as asymmetry increases. For both neutrons and protons, the cutoff dependence of the hole component is of the order of $10$ MeV and relatively independent of isospin. 

Around the Fermi surface, for all isospin asymmetries, the momentum dependence of $\Gamma_k^\tau$ is rather universal. In most cases, one cannot distinguish the curves due to different potentials within about $100$ MeV of the Fermi surface. I demonstrate again the importance of thermal effects for the $\Lambda=500$ MeV interaction at the two extreme asymmetries, $\eta=0$ (top left) and $\eta=1$ (top right panel). Whereas regions away from the Fermi surface are temperature independent, these panels show that the temperature dependence in $\Gamma_k^\tau$ is prominent around $k=k_F^\tau$. There, the quasi-particle width switches from zero at $T=0$ to a finite (and increasing) value with temperature. This is to be expected, as the behaviour around $k_F$ is governed by phase space arguments \cite{Luttinger1961,Dickhoff_book}. 

The particle $k>k_F^\tau$ component of $\Gamma^\tau_k$ is remarkably more cutoff dependent. The momentum dependence immediately above $k_F^\tau$ and up to about $500-600$ MeV is very similar in all cases, and shows an increase from zero up to an isospin-dependent maximum for the $\Lambda=450$ and $414$ MeV cases. For neutrons (protons), the peak height decreases (increases) from around $30$ MeV in symmetric matter, $\eta=0$, to around $25$ MeV ($50$ MeV) in neutron matter, $\eta=1$. For these two forces, $\Gamma_k^\tau$ becomes a decreasing function of momentum above $k \approx 500$ MeV and eventually vanishes above $k\approx 1$ 
GeV. This is stark contrast to the $\Lambda=500$ MeV case, where the peak of the particle component lies at much larger momenta in the vicinity of $k \approx 1$ GeV. The peak is also relatively independent of isospin, and very similar for neutrons and protons. The cut-off dependence clearly indicates that this component of $\Gamma_k^\tau$ depends strongly on the interaction, and the isospin dependence suggests it is mediated by isoscalar short-range physics independently of the individual component densities.

These results align with the findings in terms of spectral functions reported in Fig.~\ref{fig:sf_asy}. Whereas the width of the spectral function is qualitatively unaffected by a change in isospin asymmetry, the width of the proton spectral function changes significantly with asymmetry for hole, $k<k_F^\tau$ states. Single-particle widths provide an alternative way of characterising beyond mean-field correlations (in the Hartree-Fock approximation, $\Gamma_k=0$). The results in Fig.~\ref{fig:gamma_asy} indicate that some of its features are robust to isospin asymmetry changes, whereas others aren't. In particular, the relative strong dependence of proton single-particle widths with isospin asymmetry may be explored in terms of the isospin dependence of the absorptive component of optical potentials~\cite{Charity2006}.

\subsection{Other microscopic properties}

The microscopic results presented so far provide a very complete picture of the dynamics of infinite nuclear systems. The versatility and wide range of application is an advantage of SCGF techniques. On- and off-shell quantities provide valuable information which, when complemented with modern interactions based on $\chi$EFT, can help provide a ground base for estimates of systematic theoretical errors in simulations. 
A useful characterisation of in-medium nuclear properties can be obtained from energy-weighted moments (or sum-rules) of the spectral function. The first moment is connected to effective single-particle energies \cite{Duguet2012,Duguet2015} and has been studied in the context of both symmetric and asymmetric mater at zero and finite temperatures \cite{Polls1994,Frick2004,Rios2005c}. The conclusions reached by these studies in the early 2000s indicated that the first moment is sensitive to NN interactions, and provides a useful characterisations of correlations. The second moment has only been studied more recently~ \cite{Rios2017}. It measures fluctuations around the single-particle peak, and is a connected to the energy dependence of the imaginary part of the self-energy. Because the later is also particularly sensitive to the NN force, the second moment can provide a measure of the Hamiltonian properties in both nuclei and infinite matter. 

The SCGF method can be used to study a variety of in-medium quantities which are relevant for nuclear experiments. For instance, in the late 1990s, SCGF methods were used to provide both a formal redefinition of scattering phaseshifts in the medium  \cite{Dickhoff1998}. Practical calculations also allowed for predictions of in-medium scattering cross sections in dense matter \cite{Dickhoff1999}.The finite lifetime of quasi-particles away from the Fermi surface can be formally translated into an in-medium nucleon mean-free path. This mean-free path is relevant for a variety of nuclear physics considerations, from the validity of the shell model \cite{Negele1981} to the characterisation of multi-fragmentation reactions \cite{Lopez2014}. The consistent (via dispersion relations) description of the real and imaginary parts of the self-energy makes SCGF techniques particularly appealing in this context. In Ref.~\cite{Rios2012}, the mean-free path of nucleons in the dense infinite matter was computed by means of an extension of the SCGF formalism to the complex plane. Calculations agree well with results obtained with other methods and, most importantly, with a variety of nuclear experiments at intermediate energies \cite{Rios2012}. 

Superfluidity in nuclear systems can be detected by the appearance of in-medium pairs in the two-body scattering matrix, or the two-body propagator. This can be used to characterise the normal-superfluid phase diagram of dense matter. Extensions of the SCGF method to the superfluid domain are possible, by means of the symmetry-breaking Gorkov formalism \cite{Bozek1999,Bozek2003}. Simulations in the normal phase, however, can already provide relevant information for superfluid properties, by extrapolating normal self-energies and spectral functions to zero temperature as discussed above. With this data, one can build effective gap equations that contain the effects associated to the fragmentation of strength \cite{Muther2005}. Results with the 3 chiral interactions presented here indicate that short range correlations reduce the pairing gap, and provide a lower density gap closure compared to BCS solutions for both the singlet and the triplet channel \cite{Rios2017b}. In principle, SCGF calculations at different orders can also be used to estimate systematic uncertainties in this method. This approach may provide a better handle on the role of superfluidity at intermediate and large densities, when in particular the role of long-range correlations is poorly understood \cite{Schwenk2004,Ding2016}.

\section{Macroscopic properties}
\label{sec:macro}

\begin{figure}
\begin{center}
\includegraphics[width=0.7\linewidth]{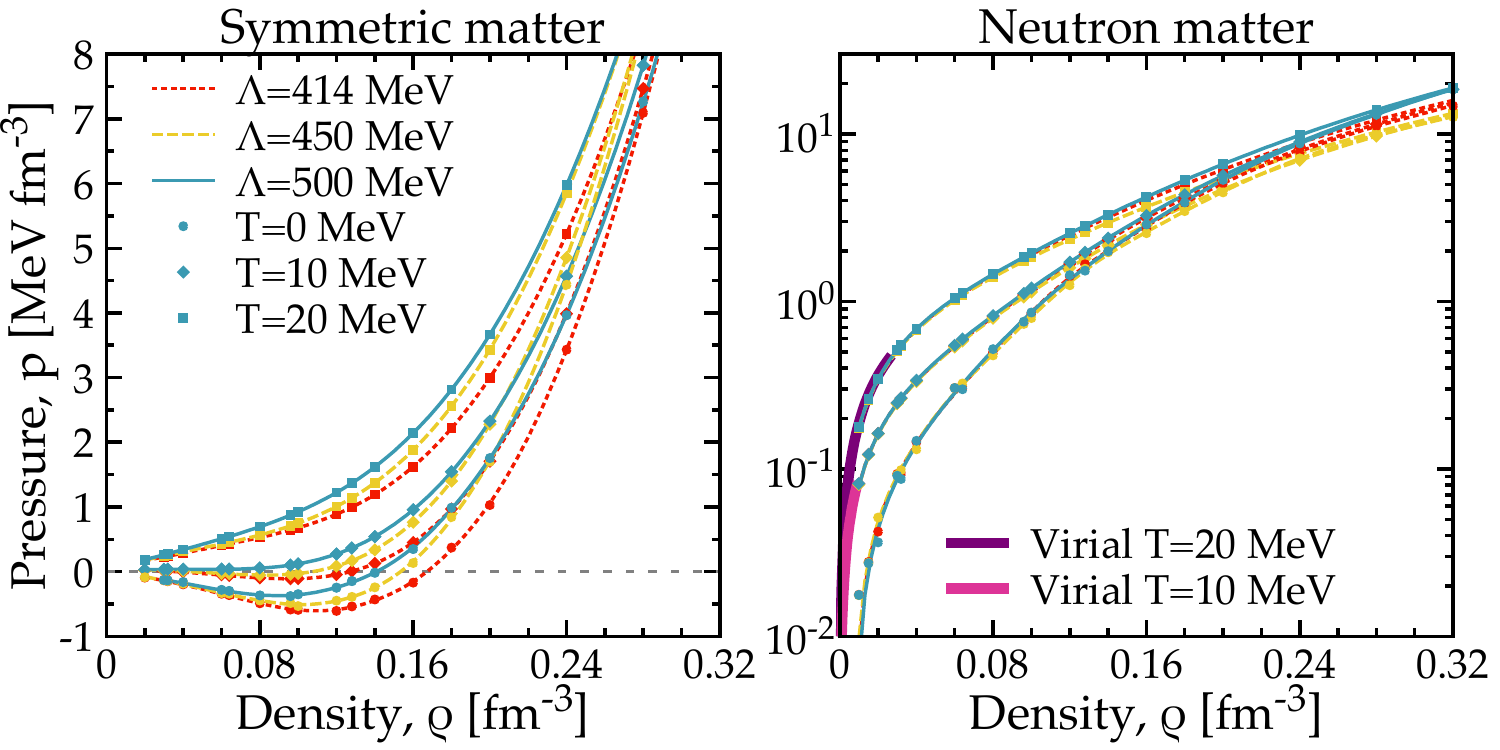}
\end{center}
\caption{
\label{fig:pressure} 
Pressure as a function of density for symmetric  (left panel) and neutron matter (right panel). Results for 3 NN interactions are shown at 3 different temperatures: $T=0$, $10$ and $20$ MeV. At low density for neutron matter, the two bands display the results from the virial expansion.
}
\end{figure}

Simulations based on SCGF methods can access the thermodynamics of the system. A wide range of other many-body methods have been extended to the finite temperature domain, including diagrammatic BHF  approaches \cite{Baldo1999}, variational methods \cite{Mukherjee2007} and perturbation theory techniques  \cite{Tolos2008,Wellenhofer2014,Wellenhofer2015}. All of these methods share common limitations in terms of the definition of quasiparticles~\cite{Wellenhofer2019}. This can lead to an inconsistency in terms of the chemical potential. On the one hand, the chemical potential is typically obtained from a normalisation condition like Eq.~(\ref{eq:dens}). On the other, one can find $\mu$ from the thermodynamical relation 
\begin{align}
\mu = \frac{\partial \epsilon}{\partial \rho} \, ,
\label{eq:muder}
\end{align}
where $\epsilon=E/V$ is the energy density. Formally and numerically, these two values may be different depending on the theoretical framework under consideration. 

A key advantage of the SCGF formulation at finite temperature is that it guarantees that the two quantities are the same, as long as the self-energy fulfils basic diagrammatic criteria \cite{Dickhoff_book}. These criteria are related to the so-called $\Phi-$derivability, and lead to a thermodynamically consistent theory \cite{Kadanoff_book}. 
In practice, this means that in SCGF calculations there is no need to perform numerically the derivative in Eq.~(\ref{eq:muder}). This is particularly convenient in calculations of the pressure, which can directly be computed from the relation $p=\rho (F/A - \mu)$. I note, however, that differences between microscopic and macroscopic chemical potentials may arise at a numerical level, particularly when the approximated treatment of 3NFs is considered \cite{Carbone2020}. 

I show the pressure as a function of density for symmetric (left panel) and neutron matter (right panel) in Fig.~\ref{fig:pressure}. These have been computed with the 3 chiral interactions discussed in this work and include the effect of 3NFs. The results are computed at two finite temperatures, $T=20$ and $10$ MeV, and are also extrapolated to the $T=0$ MeV case. Note that the extrapolation uses many more temperatures, which are not displayed here for simplicity. The discussion below is brief, and details of similar calculations comments can be found in Refs.~\cite{Rios2008,Rios2009b,Carbone2018}. 

For symmetric matter, the zero-temperature pressure shows a characteristic van der Waals shape, with a spinodal area of negative pressure that extends from zero up to the corresponding saturation density of each interaction. This structure signals the existence of a liquid-gas phase transition in nuclear matter \cite{Rios2008,Carbone2018,Baldo1999,Wellenhofer2014}. At zero temperature, the pressure vanishes at zero density and at the saturation point. Fig.~\ref{fig:pressure} indicates that the 3 chiral interactions have significantly different saturation densities, in the region $\rho_0=(0.14-0.165)$ fm$^{-3}$.
As temperature increases, the spinodal region in the pressure shrinks, until it disappears. At that stage, the pressure can still have an inflection point, which will smooth out only at the corresponding critical temperature, $T_c$, of the liquid-gas phase transition. Different many-body approaches based on the same NN interaction predict different critical points \cite{Rios2008}. Similarly, the same many-body set-up with different interactions can yield different critical points \cite{Wellenhofer2014}. 

The recent SCGF analysis presented in Ref.~\cite{Carbone2018} suggests that the critical temperature predicted by several different $\chi$EFT interactions can range from $T_c \approx 11.0$ to $T_c \approx 18$ MeV. The cutoff dependence in Fig.~\ref{fig:pressure} falls within this rage, and $T_c$ lies between about $T_c \approx 11$ MeV for the $\Lambda=500$ MeV force and $T_c \approx 16.5$ MeV for the $\Lambda=414$ MeV interaction. This relatively wide range of critical temperatures dominates over the uncertainties associated to the many-body approximation. Indeed, in Ref.~\cite{Carbone2018}, the maximum difference obtained between BHF and SCGF predictions for the critical temperature was $\approx 2$ MeV. While further many-body benchmarks may be necessary to understand the full many-body dependence of this result, it is clear that the Hamiltonian uncertainty is relevant for finite temperature predictions \cite{Carbone2020}. 
In fact, the cutoff dependence of a sub-saturation density property like $T_c$ can be taken as a sign that finite-temperature many-body correlations are significant in this region. 

The temperature dependence of the pressure in neutron matter is relevant for astrophysical considerations, particularly in the context of neutron star mergers \cite{Baiotti2017} and proto-neutron star formation \cite{Pons1999}. The right panel of Fig.~\ref{fig:pressure} shows the pressure in semilogarithmic scale as a function of density. The same three temperatures ($0, 10$ and $20$ MeV) are presented. At low densities, the pressure should be well described in terms of the model-independent virial expansion, which only depends on neutron-neutron scattering phaseshifts \cite{Horowitz2006}. The pressure for fugacities $z<0.5$ at $T=10$ and $20$ MeV is shown with two solid bands at low densities in the figure. Clearly, the low-density SCGF results agree well with virial predictions. The low-density pressure is in fact remarkably cut-off independent, and the first distinguishable cutoff-dependent features only show up above $\rho \approx 0.10$ fm$^{-3}$. This is in stark contrast to the symmetric matter case. Noticeably, this cutoff independence also holds true for the low-density, zero-temperature extrapolated results. All in all, the cutoff and many-body dependence of low-density neutron matter predictions are well under control (independently of temperature), as already discussed elsewhere \cite{Coraggio2013,Hebeler2013,Tews2013,Drischler2016b,Drischler2019}.

The cutoff dependence in neutron matter is only clearly distinguishable above saturation density. There, and towards the higher density limit of applicability of chiral interactions, the $\Lambda=500$ MeV is the interaction providing more pressure for a given density. This is followed in order of decreasing pressure by the $\Lambda=414$ MeV and the $\Lambda=450$ MeV forces. It is interesting to note that the last two switch orders with respect to symmetric matter, where for a given density the $450$ force provides more pressure than the $414$ interaction. The cutoff dependence is only one of several systematic uncertainties in these results. SCGF can also provide a footing on the many-body uncertainty, by performing calculations with self-energies obtained not only at a ladder level, but also at the first- and second-order level. The results of a recent analysis indicate that the many-body uncertainty is smaller than that associated to the underlying NN force by a factor of $2-3$ at twice saturation density \cite{Carbone2020}. 

In terms of temperature dependence, one can distinguish two distinct regimes. At sub-saturation densities, cutoff independent results dominate. At higher densities,  reaching the supra-saturation degenerate regime, the cutoff dependence overtakes temperature effects. A more quantitative analysis of the temperature dependence can be obtained by looking at the so-called thermal index (or adiabatic index) of the equation of state \cite{Constantinou2014}. This characterises the thermal dependence of the pressure (or energy) of the system. Carbone and Schwenk have recently provided an in-depth analysis of the density, temperature and hamiltonian dependence of the thermal index obtained in SCGF simulations \cite{Carbone2019}. They find that some of the assumptions made in previous literature regarding the temperature and density independence of this quantity do not hold. In addition, the effect of 3NFs is important in the high-density regime.

\begin{figure}
\begin{center}
\includegraphics[width=\linewidth]{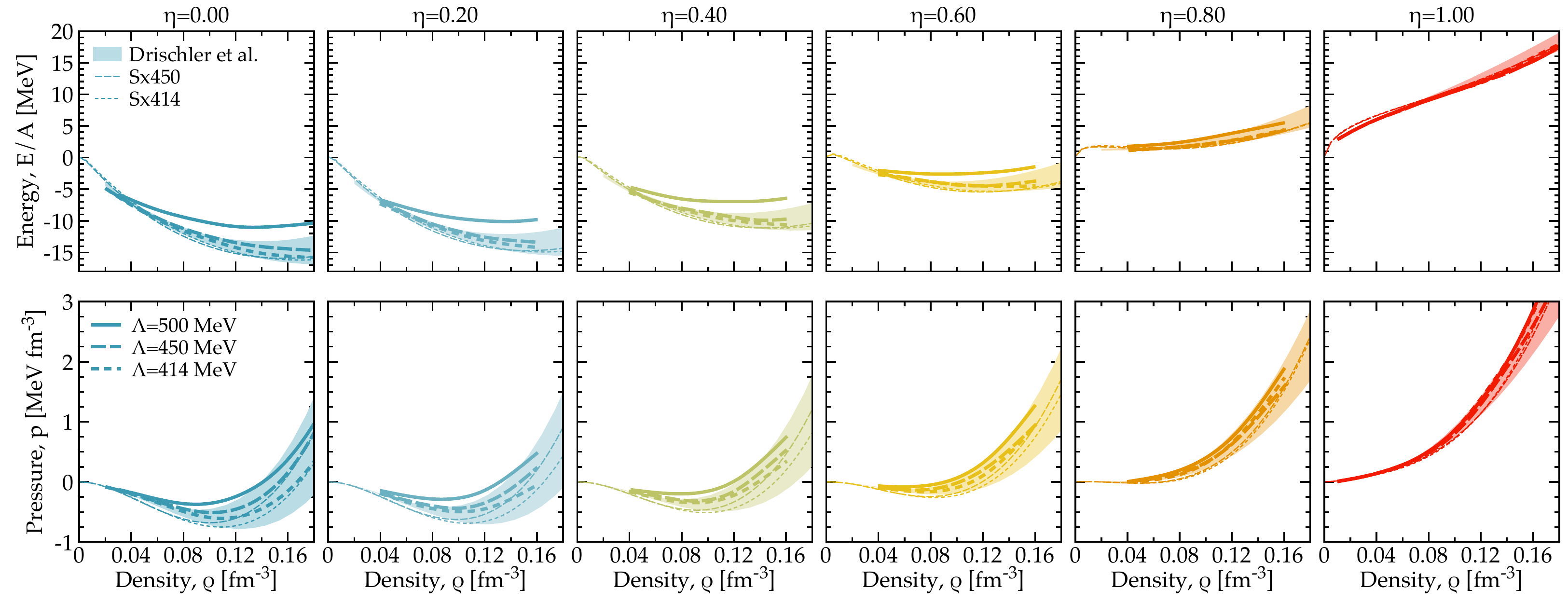}
\end{center}
\caption{
\label{fig:td_asy} 
Energy per particle (top panels) and pressure (bottom panels) as a function of density for different isospin asymmetries. Results have been extrapolated to zero temperature. The thin lines have been generated from Skyrme functionals fitted to second-order calculations with the same underlying NN forces by Lim and Holt~\cite{Lim2017}. Bands are obtained from the fits to many-body perturbation theory calculations of Drischler \emph{et al.}~\cite{Drischler2016}, 
}
\end{figure}

The two extreme limits of isospin asymmetry are interesting on their own, and provide insight on different relevant phenomena. However, both in the study of the liquid-phase transition \cite{Vidana2008} and in neutron-star astrophysics \cite{Drischler2016}, the actual isospin dependence of the results is relevant. Arbitrary isospin systems can be explored with SCGF simulations, and the results in Fig.~\ref{fig:td_asy} provide both the energy per particle (top) and the pressure (bottom panels) as a function of density for several asymmetries. Results are presented for the 3 different cutoff interactions. 

For comparison, I show two different sets of results bases on somewhat similar \emph{ab initio} calculations. The narrow lines labeled Sx450 and Sx414 correspond to the Skyrme parametrizations derived in Ref.~\cite{Lim2017}. These density functionals were in turn fit to asymmetric matter calculations performed at second order in many-body perturbation theory for the $\Lambda=450$ and $414$ chiral interactions. I do not show the corresponding Sx500 Skyrme results, which were fitted to data with  free-particle (rather than self-consistently dressed) intermediate states.
The internal dressing of fermion lines is closer in spirit to the SCGF results presented here, and one thus expects to find a better agreement. I note that if non-perturbative effects were small, one would expect an agreement between the fits and the SCGF data.

The bands in the different panels of Fig.~\ref{fig:td_asy} have been obtained from the fits to second-order many-body perturbation theory simulations for 7 different chiral interactions in Ref.~\cite{Drischler2016}. These fits reproduce the results obtained at second order in perturbation theory with internal Hartree-Fock propagators. These simulations are different to the SCGF approach in that they implement a different averaging procedure for the 3NFs, including an average value of the center-of-mass momentum. 

The results from other methods are provided here for reference and only represent a crude estimate of many-body uncertainties. The discussion below is therefore qualitative, rather than quantitative. The top left panel of Fig.~\ref{fig:td_asy} clearly indicates that all these interactions and approximations are able to saturate nuclear matter. The SCGF $\Lambda=500$ MeV results saturate at too low a density and too high an energy. In contrast, the $450$ and $414$ MeV interactions predict better saturation energies, but  are slightly off in terms of density. This may be due to deficiencies on the interactions, the many-body scheme or the implementation of the 3NFs in the SCGF approach \cite{Drischler2019}. I note however that the saturation energy itself is a poorly defined quantity and a recent reanalysis based on nuclear data suggests it may be more repulsive than anticipated \cite{Atkinson2020}. 

Comparing the energy per particle to the results of the Sx parametrizations, I find that the SCGF results for the $414$ and $450$ MeV forces  are qualitatively similar to those obtained in perturbation-theory calculations in Ref.~\cite{Coraggio2014} (for symmetric matter) and Ref.~\cite{Lim2017} (in the asymmetric matter case). The ladder resummation in the SCGF calculations seems to bring in repulsion at intermediate densities for isospin-symmetric systems. This repulsive effect, of the order of a couple of MeV in symmetric matter, reduces as the isospin asymmetry increases. 

As it is well known, the energy becomes more repulsive as $\eta$ grows. For neutron matter, and in the sub-saturation region shown in the figure, the cutoff dependence of the SCGF results is very small, certainly within $1$ MeV. The size and dependence of this cutoff dependence is in agreement with the initial findings reported in Ref.~\cite{Coraggio2013}, although the latter were obtained using third-order perturbation-theory calculations. In fact, as the isospin asymmetry of the system increases, the differences in energy between the Sx414 and Sx450 results and the SCGF calculations become much smaller. For densities above $0.06$ fm$^{-3}$, all calculations agree in the density dependence of the neutron matter energy-per particle. This is to be expected if one assumes that neutron matter is more perturbative than symmetric matter. 
The SCGF and the Sx414 and Sx450 results lie in the lower range of the band provided by Ref.~\cite{Drischler2016}, which indicates that the cutoff dependence associated to the 3 chiral interactions presented in this work is not necessarily representative of the systematic uncertainty due to different NN hamiltonians.

The reduced cutoff dependence of the results as the isospin asymmetry evolves from the left to the right panels has clear implications in terms of the symmetry energy. Because symmetric matter is more bound for the the $\Lambda=414$ and $450$ MeV forces, and the neutron matter energy is basically the same, one expects the symmetry energies of these two forces to be relatively larger than that of the $500$ MeV interaction. Indeed, computing the symmetry energy as the difference between the two extremes of isospin at $\rho=0.16$ fm$^{-3}$, one finds values that range from $S_0=31.5$ MeV for the $414$ force up to $29.8$ and $25.4$ MeV for the $450$ and $500$ MeV forces. These values are in agreement with those previously reported for these forces in Ref.~\cite{Sammarruca2015}, and are somewhat smaller than empirical determinations~\cite{Tsang2012}. 

The bottom panels of Fig.~\ref{fig:td_asy} provide a relatively similar picture in terms of the pressure. In symmetric matter (left panel), the pressure is largely negative at sub-saturation densities and, other than the $\Lambda=500$ MeV SCGF results, all results qualitatively agree. The picture evolves as isospin asymmetry increases. The spinodal, negative pressure zone decreases as $\eta$ becomes larger, as expected in general grounds \cite{Vidana2008}. For most interactions, the critical asymmetry where the spinodal area vanishes is in the range $\eta=0.6$ to $0.8$. At these relatively large asymmetries, all the different many-body predictions are contained with the bands obtained by Drischler et al. \cite{Drischler2016}. Noticeably, the $\Lambda=500$ MeV results that were off the band at small asymmetries, fall into it as neutron matter is approaches. This indicates that the discrepancies associated to this force with respect to other simulations lie in the isovector sector of the interaction. 

The cutoff dependence of the SCGF results only offers an initial indication of the size of uncertainties associated to the NN force. In qualitative terms, however, the behaviour is similar to that predicted by the bands of Ref.~\cite{Drischler2016} which, in the future, could be explored with non-perturbative methods. At the level of the interaction, there are other sources of uncertainty (like the low-energy constant or the regulator dependence) that are not really considered in this rough exploratory considerations \cite{Carbone2020}. The dependence on the many-body method is also relevant as the system becomes more and more symmetric, and chances are that quantities like the thermal index or the symmetry energy are substantially affected by the many-body method under consideration. As such, benchmark comparisons in asymmetric systems would provide an interesting testing ground for newly developed forces. 
 
\section{Conclusions}
\label{sec:conclusions}

In this work, I have briefly reviewed a set of SCGF techniques that have been used over the last couple of decades to study infinite nuclear systems. SCGF are unique in terms of their flexibility, which allows for both perturbative and non-perturbative studies. SCGF can be formulated at finite temperature and provide thermodynamically consistent results. These very same formal approach can be used to study the properties of finite nuclei with similar many-body approximations and essentially equivalent interactions \cite{Soma2020}. 
A single SCGF calculation can provide not only bulk nuclear properties, including binding energies and thermodynamics, but also a wide range of microphysics information, ranging from predictions associated to the fragmentation of single-particle strength to the characterisation of quasi-particle lifetimes. With an appropriate account of 3NFs, this method can now competitively tackle issues in infinite matter physics with the same level of predictive power as other many-body approaches. The predictions in terms of the density and isospin dependence of fragmentation of strength are relatively unique and provide an insight that is also relevant for the understanding of the evolution of beyond mean-field correlations with isospin.

In this review, I have focused on the properties of symmetric, asymmetric and neutron matter as obtained from a set of 3 $\chi$EFT interactions based on N3LO NN interactions and N2LO 3NFs. This provides a limited, but already illustrative, set of results that spans a wide range in terms of short-range physics. Of course, one would ideally perform these simulations with the widest possible set of interactions to get a more quantitative feel of the Hamiltonian dependence of the results. A more sophisticated treatment using Bayesian analysis techniques could eventually improve the estimates of systematic uncertainties \cite{Drischler2020}. 

The results indicate that the N3LO $\Lambda=500$ MeV interaction produces significantly different strength distributions than the $\Lambda=450$ and $414$ MeV forces. As expected, the $500$ MeV provide high momentum components in the momentum distribution, and also large addition energy components in the self-energy. The two softer interactions provide  results which are significantly more perturbative. In other words, their momentum distributions die out quicker in momenta and the energy dependence of the associated self-energies is much more limited. These generic results hold independently of isospin asymmetry. In fact, the results for $450$ and $414$ MeV are very similar across a wide range of density, isospins and temperatures. In neutron matter, the $500$ MeV cutoff results are also very close to the $450$ and $414$ MeV interactions. This can be translated into a strong isospin dependence in the overall cutoff dependence. This point has already been discussed in the past \cite{Coraggio2013,Sammarruca2015}, and our SCGF simulations confirm the trends from a different many-body perspective.

SCGF provide access to quasi-particle properties that are relevant in a variety of contexts. Single-particle potentials for neutrons and protons in neutron-rich matter show interesting asymmetry dependences. It is well-known that the single-particle potential for neutrons becomes less attractive with isospin asymmetry. At the same time, it becomes less sensitive to the cutoff. In contrast, the proton potential becomes more attractive in neutron-rich matter and, as asymmetry increases, the cutoff dependence is enhanced. This behaviour should be reflected in the corresponding effective masses. The asymmetry dependence of the single-particle width is also interesting, and again suggests substantial differences in the minority and majority components. 

When it comes to bulk properties, we find results that are qualitatively close to those established with perturbation-theory methods. Again, the $\Lambda=500$ MeV predictions are noticeably different than the lower cutoff results in the symmetric case. In fact, the saturation energy of this potential is about $4-5$ MeV more repulsive than empirical estimates, in contrast to the lower cutoff simulations. Results for the $450-414$ MeV forces are more consistent with each other. These also agree well with previous many-body perturbation-theory calculations based on these interactions in isospin-symmetric, isospin asymmetric and neutron matter \cite{Wellenhofer2015}. 

One can draw several conclusions from these analysis. The isospin dependence of short-range correlations can be predicted from SCGF techniques and appears to be less pronounced in bulk systems than it is in finite nuclei experiments \cite{Rios2009,Rios2014,Duer2018,Paschalis2020}. The cutoff dependence in isospin asymmetric systems affects more some single-particle properties of protons than neutrons, like their single-particle potentials. In this sense, the simulations of proton impurities in neutron matter could provide some insight into the model-dependence of neutron-proton forces. I also note that \emph{ab initio} simulations of single-particle potentials are significantly different from those predicted by phenomenological approaches \cite{Sellahewa2014}.

Looking forward, the SCGF method is at an advantage with respect to some other methods in its handling of finite temperature and isospin asymmetry effects. For saturation densities and above, this is relevant for neutron star mergers, as well as for neutron-star formation. There, the non-perturbative nature of SCGF methods can handle short-range-physics effects and hence provide meaningful results to presumably higher densities than perturbative expansions can \cite{Carbone2020}. Effects association the fragmentation of strength are also relevant in a variety of high-density astrophysical settings, where they have often been ignored in the past. These include predictions for the equation of state, but also for pairing and response properties. Broadly speaking, past SCGF simulations that have tackled these issues have not relied on systematically improvable chiral interactions that are also helpful in providing predictions for uncertainties. Interestingly, the low-density high-temperature regime of the neutrinosphere in supernova explosions is also relevant for astrophysical simulations, and can be tackled with SCGF simulations. There, SCGF methods can provide insight into
quantities like single-particle potential shifts \cite{MartinezPinedo2012,Roberts2012} and response functions needed for  neutrino processes \cite{Rrapaj2015}. Our results indicate that these should be largely interaction-independent, and the many-body dependence (if any) will provide a good handle on systematic errors. 

\section{acknowledgments}

The author thanks A. Polls, A. Carbone, J. W. Holt, C. Drischler, C. Barbieri and M. Drissi for fruitful discussion. This work is supported by the UK Science and Technology Facilities Council (STFC) Grant No. ST/P005314/1.

\bibliographystyle{iopart-num}
\bibliography{biblio}

\providecommand{\newblock}{}
\begin{thebibliography}{100}
\expandafter\ifx\csname url\endcsname\relax
  \def\url#1{{\tt #1}}\fi
\expandafter\ifx\csname urlprefix\endcsname\relax\def\urlprefix{URL }\fi
\providecommand{\eprint}[2][]{\url{#2}}

\bibitem{GW170817discovery}
Abbott B~P {\em et~al.\/} (LIGO Scientific Collaboration and Virgo
  Collaboration) 2017 {\em Phys. Rev. Lett.\/} {\bf 119}(16) 161101
  \urlprefix\url{https://link.aps.org/doi/10.1103/PhysRevLett.119.161101}

\bibitem{GW190425discovery}
Abbott B~P {\em et~al.\/} 2020 {\em ApJL\/} {\bf 892} L3
  \urlprefix\url{https://doi.org/10.3847%2F2041-8213%2Fab75f5}

\bibitem{GW170817MR1}
Abbott B~P {\em et~al.\/} (LIGO Scientific Collaboration and the Virgo
  Collaboration) 2018 {\em Phys. Rev. Lett.\/} {\bf 121}(16) 161101
  \urlprefix\url{https://link.aps.org/doi/10.1103/PhysRevLett.121.161101}

\bibitem{GW170817MR2}
Abbott B~P {\em et~al.\/} (LIGO Scientific Collaboration and Virgo
  Collaboration) 2019 {\em Phys. Rev. X\/} {\bf 9}(1) 011001
  \urlprefix\url{https://link.aps.org/doi/10.1103/PhysRevX.9.011001}

\bibitem{Raaijmakers2020}
Raaijmakers G, Greif S~K, Riley T~E, Hinderer T, Hebeler K, Schwenk A, Watts
  A~L, Nissanke S, Guillot S, Lattimer J~M and Ludlam R~M 2020 {\em ApJL\/}
  {\bf 893} L21 \urlprefix\url{https://doi.org/10.3847%2F2041-8213%2Fab822f}

\bibitem{Tsang2012}
Tsang M~B, Stone J~R, Camera F, Danielewicz P, Gandolfi S, Hebeler K, Horowitz
  C~J, Lee J, Lynch W~G, Kohley Z, Lemmon R, M\"oller P, Murakami T, Riordan S,
  Roca-Maza X, Sammarruca F, Steiner A~W, Vida\~na I and Yennello S~J 2012 {\em
  Phys. Rev. C\/} {\bf 86}(1) 015803
  \urlprefix\url{https://link.aps.org/doi/10.1103/PhysRevC.86.015803}

\bibitem{Drischler2020}
Drischler C, Furnstahl R~J, Melendez J~A and Phillips D~R 2020 How well do we
  know the neutron-matter equation of state at the densities inside neutron
  stars? {A} {B}ayesian approach with correlated uncertainties
  (\textit{Preprint} \eprint{2004.07232})
  \urlprefix\url{https://arxiv.org/abs/2004.07232}

\bibitem{Danielewicz2002}
Danielewicz P, Lacey R and Lynch W~G 2002 {\em Science\/} {\bf 298} 1592--1596
  ISSN 0036-8075
  \urlprefix\url{https://science.sciencemag.org/content/298/5598/1592}

\bibitem{Tews2020}
Tews I 2020 {\em Front. Phys.\/} {\bf 8} 153
  \urlprefix\url{https://www.frontiersin.org/article/10.3389/fphy.2020.00153}

\bibitem{Baym2018}
Baym G, Hatsuda T, Kojo T, Powell P~D, Song Y and Takatsuka T 2018 {\em Rep.
  Prog. Phys.\/} {\bf 81} 056902
  \urlprefix\url{https://doi.org/10.1088%2F1361-6633%2Faaae14}

\bibitem{Lim2018}
Lim Y and Holt J~W 2018 {\em Phys. Rev. Lett.\/} {\bf 121}(6) 062701
  \urlprefix\url{https://link.aps.org/doi/10.1103/PhysRevLett.121.062701}

\bibitem{Constantinou2014}
Constantinou C, Muccioli B, Prakash M and Lattimer J~M 2014 {\em Phys. Rev.
  C\/} {\bf 89}(6) 065802
  \urlprefix\url{https://link.aps.org/doi/10.1103/PhysRevC.89.065802}

\bibitem{Constantinou2015}
Constantinou C, Muccioli B, Prakash M and Lattimer J~M 2015 {\em Phys. Rev.
  C\/} {\bf 92}(2) 025801
  \urlprefix\url{https://link.aps.org/doi/10.1103/PhysRevC.92.025801}

\bibitem{Baiotti2017}
Baiotti L and Rezzolla L 2017 {\em Rep. Prog. Phys.\/} {\bf 80} 096901
  \urlprefix\url{https://doi.org/10.1088%2F1361-6633%2Faa67bb}

\bibitem{Duez2018}
Duez M~D and Zlochower Y 2018 {\em Rep. Prog. Phys.\/} {\bf 82} 016902
  \urlprefix\url{https://doi.org/10.1088%2F1361-6633%2Faadb16}

\bibitem{Carbone2019}
Carbone A and Schwenk A 2019 {\em Phys. Rev. C\/} {\bf 100}(2) 025805
  \urlprefix\url{https://link.aps.org/doi/10.1103/PhysRevC.100.025805}

\bibitem{Pons1999}
Pons J~A, Reddy S, Prakash M, Lattimer J~M and Miralles J~A 1999 {\em ApJ\/}
  {\bf 513} 780--804 \urlprefix\url{https://doi.org/10.1086%2F306889}

\bibitem{MartinezPinedo2012}
Mart\'{\i}nez-Pinedo G, Fischer T, Lohs A and Huther L 2012 {\em Phys. Rev.
  Lett.\/} {\bf 109}(25) 251104
  \urlprefix\url{https://link.aps.org/doi/10.1103/PhysRevLett.109.251104}

\bibitem{Rrapaj2015}
Rrapaj E, Holt J~W, Bartl A, Reddy S and Schwenk A 2015 {\em Phys. Rev. C\/}
  {\bf 91}(3) 035806
  \urlprefix\url{https://link.aps.org/doi/10.1103/PhysRevC.91.035806}

\bibitem{Gandolfi2017}
Gandolfi S, Hammer H~W, Klos P, Lynn J~E and Schwenk A 2017 {\em Phys. Rev.
  Lett.\/} {\bf 118}(23) 232501
  \urlprefix\url{https://link.aps.org/doi/10.1103/PhysRevLett.118.232501}

\bibitem{Tews2013}
Tews I, Kr\"uger T, Hebeler K and Schwenk A 2013 {\em Phys. Rev. Lett.\/} {\bf
  110}(3) 032504
  \urlprefix\url{https://link.aps.org/doi/10.1103/PhysRevLett.110.032504}

\bibitem{Epelbaum2009}
Epelbaum E, Hammer H~W and Mei\ss{}ner U~G 2009 {\em Rev. Mod. Phys.\/} {\bf
  81} 1773--1825
  \urlprefix\url{https://link.aps.org/doi/10.1103/RevModPhys.81.1773}

\bibitem{Machleidt2011}
Machleidt R and Entem D 2011 {\em Phys. Rep.\/} {\bf 503} 1 -- 75
  \urlprefix\url{http://www.sciencedirect.com/science/article/pii/S0370157311000457}

\bibitem{Ishii2007}
Ishii N, Aoki S and Hatsuda T 2007 {\em Phys. Rev. Lett.\/} {\bf 99}(2) 022001
  \urlprefix\url{https://link.aps.org/doi/10.1103/PhysRevLett.99.022001}

\bibitem{Hatsuda2018}
Hatsuda T 2018 {\em Front. Phys.\/} {\bf 13} 132105
  \urlprefix\url{https://doi.org/10.1007/s11467-018-0829-4}

\bibitem{Drischler2016b}
Drischler C, Hebeler K and Schwenk A 2016 {\em Phys. Rev. C\/} {\bf 93}(5)
  054314 \urlprefix\url{https://link.aps.org/doi/10.1103/PhysRevC.93.054314}

\bibitem{Drischler2019}
Drischler C, Hebeler K and Schwenk A 2019 {\em Phys. Rev. Lett.\/} {\bf 122}(4)
  042501
  \urlprefix\url{https://link.aps.org/doi/10.1103/PhysRevLett.122.042501}

\bibitem{Baldo2012}
Baldo M, Polls A, Rios A, Schulze H~J and Vida{\~{n}}a I 2012 {\em Phys. Rev.
  C\/} {\bf 86} 064001 ISSN 0556-2813
  \urlprefix\url{http://link.aps.org/doi/10.1103/PhysRevC.86.064001}

\bibitem{Piarulli2020}
Piarulli M, Bombaci I, Logoteta D, Lovato A and Wiringa R~B 2020 {\em Phys.
  Rev. C\/} {\bf 101}(4) 045801
  \urlprefix\url{https://link.aps.org/doi/10.1103/PhysRevC.101.045801}

\bibitem{Dickhoff2004}
Dickhoff W and Barbieri C 2004 {\em Prog. Part. Nucl. Phys.\/} {\bf 52} 377 --
  496 ISSN 0146-6410
  \urlprefix\url{http://www.sciencedirect.com/science/article/pii/S0146641004000535}

\bibitem{Dickhoff_book}
Dickhoff W~H and {Van Neck} D 2008 {\em Many-Body Theory Exposed!, 2nd
  edition\/} (New Jersey: World Scientific)

\bibitem{Barbieri2017}
Barbieri C and Carbone A 2017 {Self-Consistent Green's Function Approaches}
  {\em An Advanced Course in Computational Nuclear Physics\/} ed Hjorth-Jensen
  M, Lombardo M~P and van Kolck U (Springer) pp 571--644 ISBN 978-3-319-53336-0
  \urlprefix\url{https://doi.org/10.1007/978-3-319-53336-0_11}

\bibitem{Economou_book}
Economou E~N 2006 {\em {Green's Functions in Quantum Physics}\/} Springer
  Series in Solid-State Sciences (Springer) ISBN 3540288384

\bibitem{vanLeeuwen_book}
Stefanucci G and van Leeuwen R 2013 {\em Nonequilibrium Many-Body Theory of
  Quantum Systems\/} (Cambridge University Press) ISBN 9781139023979

\bibitem{Abrikosov1965}
Abrikosov A~A, Gorkov L~P and Dzyaloshinskii I~Y 1965 {\em {Quantum Field
  Theoretical Methods in Statistical Physics}\/} 2nd ed (Pergamon Press)

\bibitem{Mattuck_book}
Mattuck R~D 1992 {\em A guide to Feynman diagrams in the many-body problem\/}
  (Dover Publications, New York)

\bibitem{Fetter_book}
Fetter A~L and Walecka J~D 2003 {\em Quantum theory of many-particle system\/}
  1st ed (Dover)

\bibitem{Muther2000}
M{\"{u}}ther H and Polls A 2000 {\em Prog. Part. Nucl. Phys.\/} {\bf 45}
  243--334 \urlprefix\url{https://doi.org/10.1016/S0146-6410(00)00105-8}

\bibitem{Gandolfi2015}
Gandolfi S, Gezerlis A and Carlson J 2015 {\em Annu. Rev. Nucl. Part. Sci.\/}
  {\bf 65} 303--328
  \urlprefix\url{https://doi.org/10.1146/annurev-nucl-102014-021957}

\bibitem{Soma2020}
Som{\`a} V 2020 {Self-consistent Green's function theory for atomic nuclei}
  (\textit{Preprint} \eprint{2003.11321})

\bibitem{Bozek2003}
Bo{\.z}ek P 2003 {\em Phys. Lett. B\/} {\bf 551} 93
  \urlprefix\url{https://doi.org/10.1016/S0370-2693(02)03007-1}

\bibitem{Rios2009b}
Rios A, Polls A and Vida\~na I 2009 {\em Phys. Rev. C\/} {\bf 79} 025802
  \urlprefix\url{https://link.aps.org/doi/10.1103/PhysRevC.79.025802}

\bibitem{Soma2009}
Som{\`a} V and Bo{\.z}ek P 2009 {\em Phys. Rev. C\/} {\bf 80} 025803
  \urlprefix\url{https://link.aps.org/doi/10.1103/PhysRevC.80.025803}

\bibitem{Bozek1999}
Bo{\.z}ek P 1999 {\em Nucl. Phys. A\/} {\bf 657} 187
  \urlprefix\url{http://linkinghub.elsevier.com/retrieve/pii/S0375947499003255}

\bibitem{Polls1995}
Polls A, Ramos A, Gearhart C, Dickhoff W and Müther H 1995 {\em Prog. Part.
  Nucl. Phys.\/} {\bf 34} 371 -- 380

\bibitem{Rios2008}
Rios A, Polls A, Ramos A and M\"uther H 2008 {\em Phys. Rev. C\/} {\bf 78}(4)
  044314 \urlprefix\url{https://link.aps.org/doi/10.1103/PhysRevC.78.044314}

\bibitem{Carbone2018}
Carbone A, Polls A and Rios A 2018 {\em Phys. Rev. C\/} {\bf 98}(2) 025804
  \urlprefix\url{https://link.aps.org/doi/10.1103/PhysRevC.98.025804}

\bibitem{Ding2016}
Ding D, Rios A, Dussan H, Dickhoff W~H, Witte S~J, Carbone A and Polls A 2016
  {\em Phys. Rev. C\/} {\bf 94} 025802
  \urlprefix\url{https://link.aps.org/doi/10.1103/PhysRevC.94.025802}

\bibitem{Carbone2020}
Carbone A 2020 {\em Phys. Rev. Research\/} {\bf 2}(2) 023227
  \urlprefix\url{https://link.aps.org/doi/10.1103/PhysRevResearch.2.023227}

\bibitem{Jeukenne1976}
Jeukenne J, Lejeune A and Mahaux C 1976 {\em Phys. Rep.\/} {\bf 25} 83 -- 174
  \urlprefix\url{https://doi.org/10.1016/0370-1573(76)90017-X}

\bibitem{Luttinger1961}
Luttinger J~M 1961 {\em Phys. Rev.\/} {\bf 121} 942
  \urlprefix\url{https://link.aps.org/doi/10.1103/PhysRev.121.942}

\bibitem{Frick2003}
Frick T and M\"{u}ther H 2003 {\em Phys. Rev. C\/} {\bf 68} 034310
  \urlprefix\url{http://link.aps.org/doi/10.1103/PhysRevC.68.034310}

\bibitem{RiosPhD}
Rios A 2007 {\em Thermodynamical Properties of Nuclear Matter from a
  Self-Consistent Green's Function Approach,\/} Ph.D. thesis University of
  Barcelona \urlprefix\url{http://hdl.handle.net/10803/1588}

\bibitem{Cipollone2013}
Cipollone A, Barbieri C and Navr\'atil P 2013 {\em Phys. Rev. Lett.\/} {\bf
  111}(6) 062501
  \urlprefix\url{https://link.aps.org/doi/10.1103/PhysRevLett.111.062501}

\bibitem{Carbone2013a}
Carbone A, Cipollone A, Barbieri C, Rios A and Polls A 2013 {\em Phys. Rev.
  C\/} {\bf 88} 054326
  \urlprefix\url{http://link.aps.org/doi/10.1103/PhysRevC.88.054326}

\bibitem{Rios2006}
Rios A, Polls A, Ramos A and M\"uther H 2006 {\em Phys. Rev. C\/} {\bf 74}(5)
  054317 \urlprefix\url{https://link.aps.org/doi/10.1103/PhysRevC.74.054317}

\bibitem{Pethick1973}
Pethick C~J and Carneiro G~M 1973 {\em Phys. Rev. A\/} {\bf 7}(1) 304--318
  \urlprefix\url{https://link.aps.org/doi/10.1103/PhysRevA.7.304}

\bibitem{Carneiro1975}
Carneiro G~M and Pethick C~J 1975 {\em Phys. Rev. B\/} {\bf 11}(3) 1106--1124
  \urlprefix\url{https://link.aps.org/doi/10.1103/PhysRevB.11.1106}

\bibitem{Kadanoff_book}
Kadanoff L and Baym G 1962 {\em Quantum statistical mechanics\/} (W.A.
  Benjamin)

\bibitem{Rios2005}
Frick T, M\"uther H, Rios A, Polls A and Ramos A 2005 {\em Phys. Rev. C\/} {\bf
  71}(1) 014313
  \urlprefix\url{https://link.aps.org/doi/10.1103/PhysRevC.71.014313}

\bibitem{Konrad2005}
Konrad P, Lenske H and Mosel U 2005 {\em Nucl. Phys. A\/} {\bf 756} 192 -- 212
  \urlprefix\url{https://doi.org/10.1016/j.nuclphysa.2005.03.083}

\bibitem{Rios2009}
Rios A, Polls A and Dickhoff W~H 2009 {\em Phys. Rev. C\/} {\bf 79} 064308
  \urlprefix\url{https://link.aps.org/doi/10.1103/PhysRevC.79.064308}

\bibitem{Rios2014}
Rios A, Polls A and Dickhoff W~H 2014 {\em Phys. Rev. C\/} {\bf 89} 044303
  \urlprefix\url{https://link.aps.org/doi/10.1103/PhysRevC.89.044303}

\bibitem{Carbone2014}
Carbone A, Rios A and Polls A 2014 {\em Phys. Rev. C\/} {\bf 90} 054322
  \urlprefix\url{https://link.aps.org/doi/10.1103/PhysRevC.90.054322}

\bibitem{Wellenhofer2015}
Wellenhofer C, Holt J~W and Kaiser N 2015 {\em Phys. Rev. C\/} {\bf 92}(1)
  015801 \urlprefix\url{https://link.aps.org/doi/10.1103/PhysRevC.92.015801}

\bibitem{Coraggio2013}
Coraggio L, Holt J~W, Itaco N, Machleidt R and Sammarruca F 2013 {\em Phys.
  Rev. C\/} {\bf 87}(1) 014322
  \urlprefix\url{https://link.aps.org/doi/10.1103/PhysRevC.87.014322}

\bibitem{Coraggio2014}
Coraggio L, Holt J~W, Itaco N, Machleidt R, Marcucci L~E and Sammarruca F 2014
  {\em Phys. Rev. C\/} {\bf 89}(4) 044321
  \urlprefix\url{https://link.aps.org/doi/10.1103/PhysRevC.89.044321}

\bibitem{Holt2010}
Holt J~W, Kaiser N and Weise W 2010 {\em Phys. Rev. C\/} {\bf 81} 024002 ISSN
  0556-2813 \urlprefix\url{http://link.aps.org/doi/10.1103/PhysRevC.81.024002}

\bibitem{Holt2020}
Holt J~W, Kawaguchi M and Kaiser N 2020 {\em Front. Phys.\/} {\bf 8} 100
  \urlprefix\url{https://www.frontiersin.org/article/10.3389/fphy.2020.00100}

\bibitem{Carbone2013}
Carbone A, Polls A and Rios A 2013 {\em Phys. Rev. C\/} {\bf 88} 044302
  \urlprefix\url{https://link.aps.org/doi/10.1103/PhysRevC.88.044302}

\bibitem{Drischler2016}
Drischler C, Carbone A, Hebeler K and Schwenk A 2016 {\em Phys. Rev. C\/} {\bf
  94}(5) 054307
  \urlprefix\url{https://link.aps.org/doi/10.1103/PhysRevC.94.054307}

\bibitem{Rios2017b}
Rios A, Polls A and Dickhoff W~H 2017 {\em J. Low T. Phys.\/} {\bf 189}
  234--249 \urlprefix\url{https://doi.org/10.1007/s10909-017-1818-7}

\bibitem{Ramos1991}
Ramos A, Dickhoff W~H and Polls A 1991 {\em Phys. Rev. C\/} {\bf 43} 2239--2253
  \urlprefix\url{https://link.aps.org/doi/10.1103/PhysRevC.43.2239}

\bibitem{Rios2017}
Rios A, Carbone A and Polls A 2017 {\em Phys. Rev. C\/} {\bf 96}(1) 014003
  \urlprefix\url{https://link.aps.org/doi/10.1103/PhysRevC.96.014003}

\bibitem{Rohe2004}
Rohe D {\em et~al.\/} (E97-006 Collaboration) 2004 {\em Phys. Rev. Lett.\/}
  {\bf 93}(18) 182501
  \urlprefix\url{https://link.aps.org/doi/10.1103/PhysRevLett.93.182501}

\bibitem{Schmidt2020}
Schmidt A {\em et~al.\/} (CLAS Collaboration) 2020 {\em Nature\/} {\bf 578}
  540--544 \urlprefix\url{https://doi.org/10.1038/s41586-020-2021-6}

\bibitem{TypelBrown2001}
Typel S and Brown B~A 2001 {\em Phys. Rev. C\/} {\bf 64}(2) 027302
  \urlprefix\url{https://link.aps.org/doi/10.1103/PhysRevC.64.027302}

\bibitem{Haensel}
Haensel P, Potekhin A~Y and Yakovlev D~G 2007 {\em Neutron Stars 1: Equation of
  State and Structure\/} (Springer) ISBN 0-387-33543-9

\bibitem{Tang2003}
Tang A {\em et~al.\/} 2003 {\em Phys. Rev. Lett.\/} {\bf 90} 042301
  \urlprefix\url{https://link.aps.org/doi/10.1103/PhysRevLett.90.042301}

\bibitem{Piasetzky2006}
Piasetzky E, Sargsian M, Frankfurt L, Strikman M and Watson J~W 2006 {\em Phys.
  Rev. Lett.\/} {\bf 97} 162504
  \urlprefix\url{https://link.aps.org/doi/10.1103/PhysRevLett.97.162504}

\bibitem{Hen2014}
Hen O, Sargsian M, Weinstein L~B, Piasetzky E {\em et~al.\/} 2014 {\em
  Science\/} {\bf 346} 614--7
  \urlprefix\url{https://doi.org/10.1126/science.1256785}

\bibitem{Sargsian2014}
Sargsian M~M 2014 {\em Phys. Rev. C\/} {\bf 89} 034305
  \urlprefix\url{https://link.aps.org/doi/10.1103/PhysRevC.89.034305}

\bibitem{Duer2018}
Duer M, Hen O, Piasetzky E {\em et~al.\/} (CLAS Collaboration) 2018 {\em
  Nature\/} {\bf 560} 617--621
  \urlprefix\url{https://doi.org/10.1038/s41586-018-0400-z}

\bibitem{Ryckebusch2019}
Ryckebusch J, Cosyn W, Stevens S, Casert C and Nys J 2019 {\em Phys. Lett. B\/}
  {\bf 792} 21 -- 28
  \urlprefix\url{https://doi.org/10.1016/j.physletb.2019.03.016}

\bibitem{Paschalis2020}
Paschalis S, Petri M, Macchiavelli A, Hen O and Piasetzky E 2020 {\em Phys.
  Lett. B\/} {\bf 800} 135110
  \urlprefix\url{https://doi.org/10.1016/j.physletb.2019.135110}

\bibitem{Carbone2012}
Carbone A, Polls A and Rios A 2012 {\em {EPL} (Europhysics Letters)\/} {\bf 97}
  22001 \urlprefix\url{https://doi.org/10.1209%2F0295-5075%2F97%2F22001}

\bibitem{Hen2015}
Hen O, Li B~A, Guo W~J, Weinstein L~B and Piasetzky E 2015 {\em Phys. Rev. C\/}
  {\bf 91}(2) 025803
  \urlprefix\url{https://link.aps.org/doi/10.1103/PhysRevC.91.025803}

\bibitem{Cai2016}
Cai B~J and Li B~A 2016 {\em Phys. Rev. C\/} {\bf 93} 014619
  \urlprefix\url{https://link.aps.org/doi/10.1103/PhysRevC.93.014619}

\bibitem{Gade2008}
Gade A {\em et~al.\/} 2008 {\em Phys. Rev. C\/} {\bf 77}(4) 044306
  \urlprefix\url{https://link.aps.org/doi/10.1103/PhysRevC.77.044306}

\bibitem{Roberts2012}
Roberts L~F, Reddy S and Shen G 2012 {\em Phys. Rev. C\/} {\bf 86}(6) 065803
  \urlprefix\url{https://link.aps.org/doi/10.1103/PhysRevC.86.065803}

\bibitem{Zuo2002}
Zuo W, Lejeune A, Lombardo U and Mathiot J~F 2002 {\em Eur. Phys. J. A.\/} {\bf
  14} 469--475 \urlprefix\url{https://doi.org/10.1140/epja/i2002-10031-y}

\bibitem{Zuo2005}
Zuo W, Cao L~G, Li B~A, Lombardo U and Shen C~W 2005 {\em Phys. Rev. C\/} {\bf
  72}(1) 014005
  \urlprefix\url{https://link.aps.org/doi/10.1103/PhysRevC.72.014005}

\bibitem{Holt2016}
Holt J~W, Kaiser N and Miller G~A 2016 {\em Phys. Rev. C\/} {\bf 93}(6) 064603
  \urlprefix\url{https://link.aps.org/doi/10.1103/PhysRevC.93.064603}

\bibitem{Sellahewa2014}
Sellahewa R and Rios A 2014 {\em Phys. Rev. C\/} {\bf 90}(5) 054327
  \urlprefix\url{https://link.aps.org/doi/10.1103/PhysRevC.90.054327}

\bibitem{Hebeler2013}
Hebeler K, Lattimer J~M, Pethick C~J and Schwenk A 2013 {\em ApJ\/} {\bf 773}
  11 \urlprefix\url{https://doi.org/10.1088%2F0004-637x%2F773%2F1%2F11}

\bibitem{Drischler2014}
Drischler C, Som\`a V and Schwenk A 2014 {\em Phys. Rev. C\/} {\bf 89}(2)
  025806 \urlprefix\url{https://link.aps.org/doi/10.1103/PhysRevC.89.025806}

\bibitem{Charity2006}
Charity R~J, Sobotka L~G and Dickhoff W~H 2006 {\em Phys. Rev. Lett.\/} {\bf
  97}(16) 162503
  \urlprefix\url{https://link.aps.org/doi/10.1103/PhysRevLett.97.162503}

\bibitem{Duguet2012}
Duguet T and Hagen G 2012 {\em Phys. Rev. C\/} {\bf 85} 034330
  \urlprefix\url{http://link.aps.org/doi/10.1103/PhysRevC.85.034330}

\bibitem{Duguet2015}
Duguet T, Hergert H, Holt J~D and Som{\`{a}} V 2015 {\em Phys. Rev. C\/} {\bf
  92} 034313 \urlprefix\url{http://link.aps.org/doi/10.1103/PhysRevC.92.034313}

\bibitem{Polls1994}
Polls A, Ramos A, Ventura J, Amari S and Dickhoff W~H 1994 {\em Phys. Rev. C\/}
  {\bf 49} 3050--3054
  \urlprefix\url{http://link.aps.org/doi/10.1103/PhysRevC.49.3050}

\bibitem{Frick2004}
Frick T, M{\"{u}}ther H and Polls A 2004 {\em Phys. Rev. C\/} {\bf 69} 054305
  \urlprefix\url{http://link.aps.org/doi/10.1103/PhysRevC.69.054305}

\bibitem{Rios2005c}
Rios A, Polls A and M{\"{u}}ther H 2006 {\em Phys. Rev. C\/} {\bf 73} 024305
  \urlprefix\url{https://link.aps.org/doi/10.1103/PhysRevC.73.024305}

\bibitem{Dickhoff1998}
Dickhoff W~H 1998 {\em Phys. Rev. C\/} {\bf 58}(5) 2807--2820
  \urlprefix\url{https://link.aps.org/doi/10.1103/PhysRevC.58.2807}

\bibitem{Dickhoff1999}
Dickhoff W~H, Gearhart C~C, Roth E~P, Polls A and Ramos A 1999 {\em Phys. Rev.
  C\/} {\bf 60}(6) 064319
  \urlprefix\url{https://link.aps.org/doi/10.1103/PhysRevC.60.064319}

\bibitem{Negele1981}
Negele J and Yazaki K 1981 {\em Phys. Rev. Lett.\/} {\bf 47} 71--74
  \urlprefix\url{http://link.aps.org/doi/10.1103/PhysRevLett.47.71}

\bibitem{Lopez2014}
Lopez O {\em et~al.\/} (INDRA Collaboration) 2014 {\em Phys. Rev. C\/} {\bf
  90}(6) 064602
  \urlprefix\url{https://link.aps.org/doi/10.1103/PhysRevC.90.064602}

\bibitem{Rios2012}
Rios A and Som{\`{a}} V 2012 {\em Phys. Rev. Lett.\/} {\bf 108} 012501
  \urlprefix\url{http://link.aps.org/doi/10.1103/PhysRevLett.108.012501}

\bibitem{Muther2005}
M{\"{u}}ther H and Dickhoff W~H 2005 {\em Phys. Rev. C\/} {\bf 72} 054313
  \urlprefix\url{http://link.aps.org/doi/10.1103/PhysRevC.72.054313}

\bibitem{Schwenk2004}
Schwenk A and Friman B 2004 {\em Phys. Rev. Lett.\/} {\bf 92}(8) 082501
  \urlprefix\url{https://link.aps.org/doi/10.1103/PhysRevLett.92.082501}

\bibitem{Baldo1999}
Baldo M and Ferreira L 1999 {\em Phys. Rev. C\/} {\bf 59} 682--703
  \urlprefix\url{http://link.aps.org/doi/10.1103/PhysRevC.59.682}

\bibitem{Mukherjee2007}
Mukherjee A and Pandharipande V~R 2007 {\em Phys. Rev. C\/} {\bf 75}(3) 035802
  \urlprefix\url{https://link.aps.org/doi/10.1103/PhysRevC.75.035802}

\bibitem{Tolos2008}
Tolos L, Friman B and Schwenk A 2008 {\em Nucl. Phys. A\/} {\bf 806} 105--116
  \urlprefix\url{https://linkinghub.elsevier.com/retrieve/pii/S0375947408004077}

\bibitem{Wellenhofer2014}
Wellenhofer C, Holt J~W, Kaiser N and Weise W 2014 {\em Phys. Rev. C\/} {\bf
  89}(6) 064009
  \urlprefix\url{https://link.aps.org/doi/10.1103/PhysRevC.89.064009}

\bibitem{Wellenhofer2019}
Wellenhofer C 2019 {\em Phys. Rev. C\/} {\bf 99}(6) 065811
  \urlprefix\url{https://link.aps.org/doi/10.1103/PhysRevC.99.065811}

\bibitem{Horowitz2006}
Horowitz C and Schwenk A 2006 {\em Phys. Lett. B\/} {\bf 638} 153--159
  \urlprefix\url{https://linkinghub.elsevier.com/retrieve/pii/S0370269306006599}

\bibitem{Lim2017}
Lim Y and Holt J~W 2017 {\em Phys. Rev. C\/} {\bf 95}(6) 065805
  \urlprefix\url{https://link.aps.org/doi/10.1103/PhysRevC.95.065805}

\bibitem{Vidana2008}
Vida{\~n}a I and Polls A 2008 {\em Phys. Lett. B\/} {\bf 666} 232 -- 238
  \urlprefix\url{http://www.sciencedirect.com/science/article/pii/S0370269308008848}

\bibitem{Atkinson2020}
Atkinson M~C, Dickhoff W~H, Piarulli M, Rios A and Wiringa R~B 2020 Revisiting
  the relation between the binding energy of finite nuclei and the equation of
  state of infinite nuclear matter (\textit{Preprint}
  \eprint{arxiv:2001.07231}) \urlprefix\url{https://arxiv.org/abs/2001.07231}

\bibitem{Sammarruca2015}
Sammarruca F, Coraggio L, Holt J~W, Itaco N, Machleidt R and Marcucci L~E 2015
  {\em Phys. Rev. C\/} {\bf 91}(5) 054311
  \urlprefix\url{https://link.aps.org/doi/10.1103/PhysRevC.91.054311}

\end{thebibliography}

\end{document}